\documentclass[a4paper,fleqn,usenatbib]{mnras}

\usepackage{newtxtext,newtxmath}

\usepackage[T1]{fontenc}
\usepackage{ae,aecompl}

\usepackage[dvipdfmx]{graphicx}	
\usepackage{amsmath}	
\usepackage{amssymb}	

\usepackage{lscape}
\usepackage{subfigure}
\usepackage{multirow}
\usepackage{url}
\usepackage{threeparttable}
\usepackage{color}
\usepackage{natbib, aas_macros}
\title[NIR Spectroscopy of P Cygni]{A newly-identified emission-line region around P Cygni}

\author[M.\,Mizumoto et al.]{Misaki Mizumoto$^{1,2}$\thanks{E-mail: misaki.mizumoto@durham.ac.uk, mizumoto.misaki@gmail.com (MM)},
Naoto Kobayashi$^{3,4,5}$,
Satoshi Hamano$^{5}$, 
Yuji Ikeda$^{5,6}$, 
\newauthor
Sohei Kondo$^{5}$, 
Hiroaki Sameshima$^{5}$,
Noriyuki Matsunaga$^{1,5}$, 
Kei Fukue$^{5}$, 
\newauthor
Chikako Yasui$^{7}$, 
Natsuko Izumi$^{7}$,
Hideyo Kawakita$^{5,8}$,
Kenshi Nakanishi$^{5,8}$,
\newauthor
Tetsuya Nakaoka$^{8,5}$,
Shogo Otsubo$^{8,5}$
\& Hiroyuki Maehara$^{9}$
\\
$^1$Department of Astronomy, Graduate School of Science, The University of Tokyo, Hongo, Bunkyo-ku, Tokyo, 113-0033, Japan\\
$^2$Centre for Extragalactic Astronomy, Department of Physics, University of Durham, South Road, Durham, DH1 3LE, UK\\
$^3$Kiso Observatory, Institute of Astronomy, School of Science, The University of Tokyo, Mitake, Kiso-machi, Kiso-gun,\\
 Nagano, 397-0101, Japan\\
$^4$Institute of Astronomy, Graduate School of Science, The University of Tokyo, Osawa, Mitaka, Tokyo, 181-0015, Japan\\
$^5$Laboratory of IR High-resolution Spectroscopy (LIH), Koyama Astronomical Observatory, Kyoto Sangyo University, Motoyama,\\
Kamigamo, Kita-Ku, Kyoto, 603-8555, Japan\\
$^6$Photocoding, Iwakura-Kita-Ikedacho, Sakyo-Ku, Kyoto, 606-0004, Japan\\
$^7$National Astronomical Observatory of Japan, Osawa, Mitaka, Tokyo 181-8588, Japan\\
$^8$Department of Physics, Faculty of Science, Kyoto Sangyo University, Motoyama, Kamigamo, Kita-ku, Kyoto, 603-8555, Japan\\
$^9$Okayama Astrophysical Observatory, National Astronomical Observatory of Japan, Honjo, Kamogata, Asakuchi, Okayama,\\
 719-0232, Japan
}

\date{Accepted XXX. Received YYY; in original form ZZZ}

\pubyear{2017}

\begin{document}
\label{firstpage}
\pagerange{\pageref{firstpage}--\pageref{lastpage}}
\maketitle


\begin{abstract}
We present a high-resolution ($R \simeq 20,000$) near-infrared (9,100--13,500~\AA) long-slit spectrum of P Cygni obtained with the newly commissioned WINERED spectrograph in Japan. 
In the obtained spectrum, we have found that
the velocity profiles of the [\ion{Fe}{II}] emission lines are resolved into two peaks at a velocity of $\simeq220$~km~s$^{-1}$ with a moderate dip in between and with additional sub-peaks at $\simeq\pm100$~km~s$^{-1}$.
The sub-peak component is confirmed with the long-slit echellogram to originate in the known shell with a radius of $\simeq10^{\prime\prime}$,
which was originally created by the outburst in 1600 AD.
On the other hand, the $\simeq220$~km~s$^{-1}$ component, 
which dominates the [\ion{Fe}{II}] flux from P Cygni, 
is found to be concentrated closer to the central star with an apparent spatial extent of $\simeq3^{\prime\prime}$.
The extent is much larger than the compact ($<0^{\prime\prime}.1$) regions traced with hydrogen, helium, and metal permitted lines.
The velocity, estimated mass, and dynamical time of the extended emission-line region suggest that
the region is an outer part of the stellar wind region.
We suggest that the newly-identified emission-line region may trace a reverse shock due to the stellar wind overtaking the outburst shell.
\end{abstract}



\section{Introduction}

Luminous blue variable (LBV) stars are believed to have evolved from high-mass main-sequence stars and to become eventually Wolf-Rayet stars after various magnitudes of mass-loss (e.g., \citealt{lan94,mey11,smi14} and references therein).
Circumstellar nebulae around LBV stars provide essential clues for mass-loss events (e.g., \citealt{naj97}) and 
evolutionary process of LBV stars (e.g., \citealt{smi11,wei11}).
Steady mass-loss events create emission photospheres or shells around stars (e.g., \citealt{lam85}),
whereas flash mass-loss events, which are often called ``eruptions'' or ``outbursts'', 
create emission shells as those seen in P Cygni and $\eta$ Carinae (e.g., \citealt{lam86,hum99,ish03,smihar06}).
Thus the structure and kinematics of the LBV nebulae give us important clues about the mass-loss events.

Infrared (IR) [\ion{Fe}{II}] lines are a powerful tool
for investigating the structure and physical mechanisms of the LBV mass-loss events \citep{smi02}.
Bright IR [\ion{Fe}{II}] lines are typically detected toward shock-excited objects or outflowing objects, such as supernova remnants, massive stars, and low-mass young stellar objects (e.g., \citealt{mea00,smi01a,smi02,har04,lee09,shi13}), and
are detected in all the observed LBVs with near-IR (NIR) spectroscopy \citep{smi02}.
Because the LBV nebulae are optically thin at the IR wavelength region,
IR observations are key tools to study the circumstellar nebulae of LBVs (e.g., \citealt{art11,cla11,gva12}).
For example, NIR observations of $\eta$ Carinae clearly resolve the emission components from different areas \citep{smi02c,smi06}.

\citet{bar94} found a circular nebula around P Cygni with an angular radius of $\sim11^{\prime\prime}$ and an expansion velocity of $140$~km~s$^{-1}$ using the optical [\ion{N}{II}] $\lambda$6584 line.
\citet{smihar06} (hereafter SH06) deeply investigated this nebula using the NIR [\ion{Fe}{II}] $\lambda$16440 line,
and concluded that this nebula with a shell-like structure with a radius of $\sim8^{\prime\prime}-10^{\prime\prime}$ was ejected by the 1600-AD outburst.
Other circumstellar nebulae outside of the 1600-AD outburst shell have also been studied (e.g., \citealt{mea96,mea00}).
In addition, the [\ion{Fe}{II}] emission of P Cygni inside the 1600-AD outburst shell has also been investigated and
 found to be created in the region of $R\gtrsim100R_*$, where $R_*$ is the radius of P Cygni ($=76\pm14\,R_\odot$; \citealt{lam83}), based on the line ratio and/or the velocity width of the optical [\ion{Fe}{II}] emission lines (e.g., \citealt{isr91,sta91,mar97}).
\citet{mar00} suggested that the forbidden lines in P Cygni originate at $\simeq100R_*$, where the wind has reached its terminal velocity.
However, to our surprise, the upper limit for the radius of the [\ion{Fe}{II}] emission region has not been explicitly constrained.
\citet{arc14} shows the [\ion{Fe}{II}] $\lambda16440$ image of P Cygni nebula with high angular resolution by adaptive-optics observations, but
the inner structure within $3^{\prime\prime}$ is unclear because they masked the region.

Here, we observed P Cygni with a newly developed high-resolution ($R\simeq20,000$) NIR ($9,100$--$13,500$~\AA) echelle ``WINERED'' spectrograph.
P Cygni is the nearest LBV star at a distance of 1.7~kpc \citep{naj97} and thus the best target to study mass-loss events in LBVs.
The range of the studied wavelengths, 9,100--13,500~\AA, falls between the optical and infrared wavelengths and has been relatively poorly studied in astronomy. 
However, this ``niche'' wavelength-range is rich in various types of emission lines, yet is contaminated less with background lines than optical wavelengths.

In this paper, we first describe the WINERED spectrograph and the observation in Section 2.
In Section 3, we show that the velocity profiles of the NIR [\ion{Fe}{II}] lines have a clear ``double-peak'', which have been considered as ``flat-topped'' in the past studies, and 
constrain the spatial extent of the [\ion{Fe}{II}] emission region.
Finally, we discuss the origin of the extended region of the [\ion{Fe}{II}] emission in Section 4.


\section{Observations and data reduction}

\begin{figure}
\centering
\includegraphics[width=90mm]{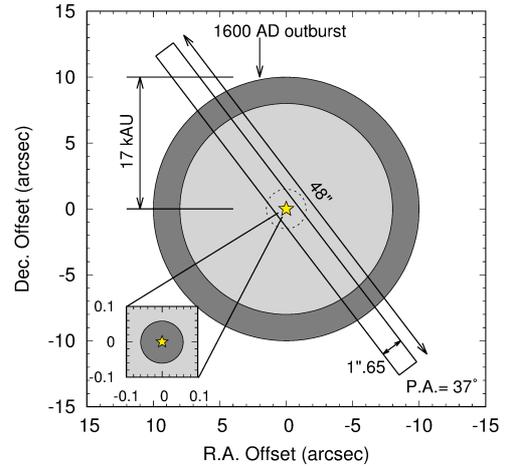} 
\caption{The position and P.A.\ of the slit for the WINERED observation of P Cygni and its nebula.
The horizontal and vertical axes show the offsets in the right ascension and declination, respectively, in arcsec from the position of P Cygni, 
which is denoted as the star symbol at the centre. 
The tilted and slender rectangle shows the position of the slit on the sky.
The dark-grey ring with radii of 8$^{\prime\prime}$--10$^{\prime\prime}$ illustrates the 1600-AD outburst shell (SH06).
The dark-grey filled circle with a radius of $\sim 0^{\prime\prime}.06$ in the inset illustrates the extent of the steady mass-loss wind region ($\sim100$ AU).
The dotted line is a circle with a diameter of $2^{\prime\prime}.9$ ($=4.9$~kAU), which demonstrates the extent of the [\ion{Fe}{II}] emission-line region we have discovered (see the main text in \S\ref{sec:spatial}).
}
\label{fig1}
\end{figure}

We obtained the spectrum of P Cygni on 2014 September 20 with the NIR high-resolution spectrograph WINERED \citep{ike16},
mounted on the F/10 Nasmyth focus of the Araki 1.3 m telescope at Koyama Astronomical Observatory, Japan \citep{yos12}.
WINERED uses a 1.7 $\mathrm{\mu m}$ cut-off $2048\times2048$ HAWAII-2RG array 
with a pixel scale of 0$^{\prime\prime}$.8 pixel$^{-1}$,
covering the wavelength range of $9,100$--$13,500$~\AA\ simultaneously. 
Fig.\ \ref{fig1} shows the slit position relative to the central star of P Cygni. 
We used a long slit with a length of 48$^{\prime\prime}$, oriented along P.A.~$\simeq 37^\circ$.
The slit width is 1$^{\prime\prime}$.6 (2 pix),
 which provides the maximum spectral resolving power of $R\simeq28,300$ or $\Delta v\simeq11$~km~s$^{-1}$.
However, our observed spectrum has a lower spectral resolution of $R\simeq20,000$ ($\Delta v\simeq15$~km~s$^{-1}$) due to the optical setting during the observing period.
The exposure time was 150 seconds per frame, which was long enough to make some strong \ion{H}{I} and \ion{He}{I} lines saturated in the obtained spectra.
We observed P Cygni 8 times and combined all the frames; thus the total exposure time was 1200 sec.
The one-dimensional spectrum was created, using the data within the slit-length of $\simeq10^{\prime\prime}$.
The obtained S/N ratio was $130$.
The wavelength was calibrated using the spectrum of a Th-Ar lamp in front of the slit.
The uncertainty of the central velocities of the lines was smaller than 1~km~s$^{-1}$.
For the telluric-absorption correction,
we used the spectrum of the spectrometric standard star 28 Cet,
obtained at the similar airmass to that in our observations. 
The sky conditions were photometric and the seeing was relatively good for the site (FWHM~$\sim 4^{\prime\prime}$).

\section{Results}
\subsection{Velocity profiles}\label{result}

\begin{figure*}
  \begin{center}
\subfigure{
\resizebox{7cm}{!}{\includegraphics[width=45mm,angle=270]{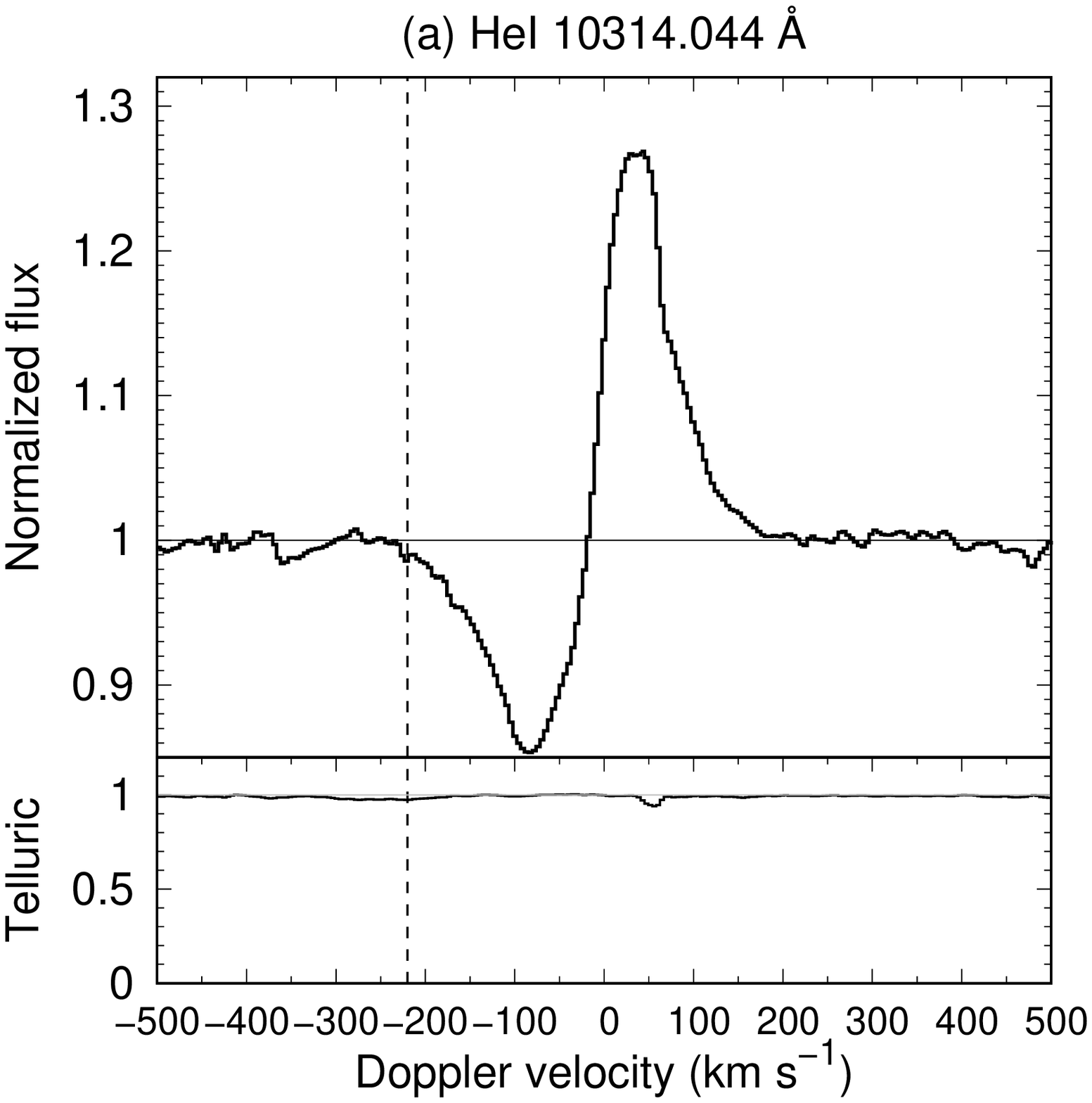}}
\resizebox{7cm}{!}{\includegraphics[width=45mm,angle=270]{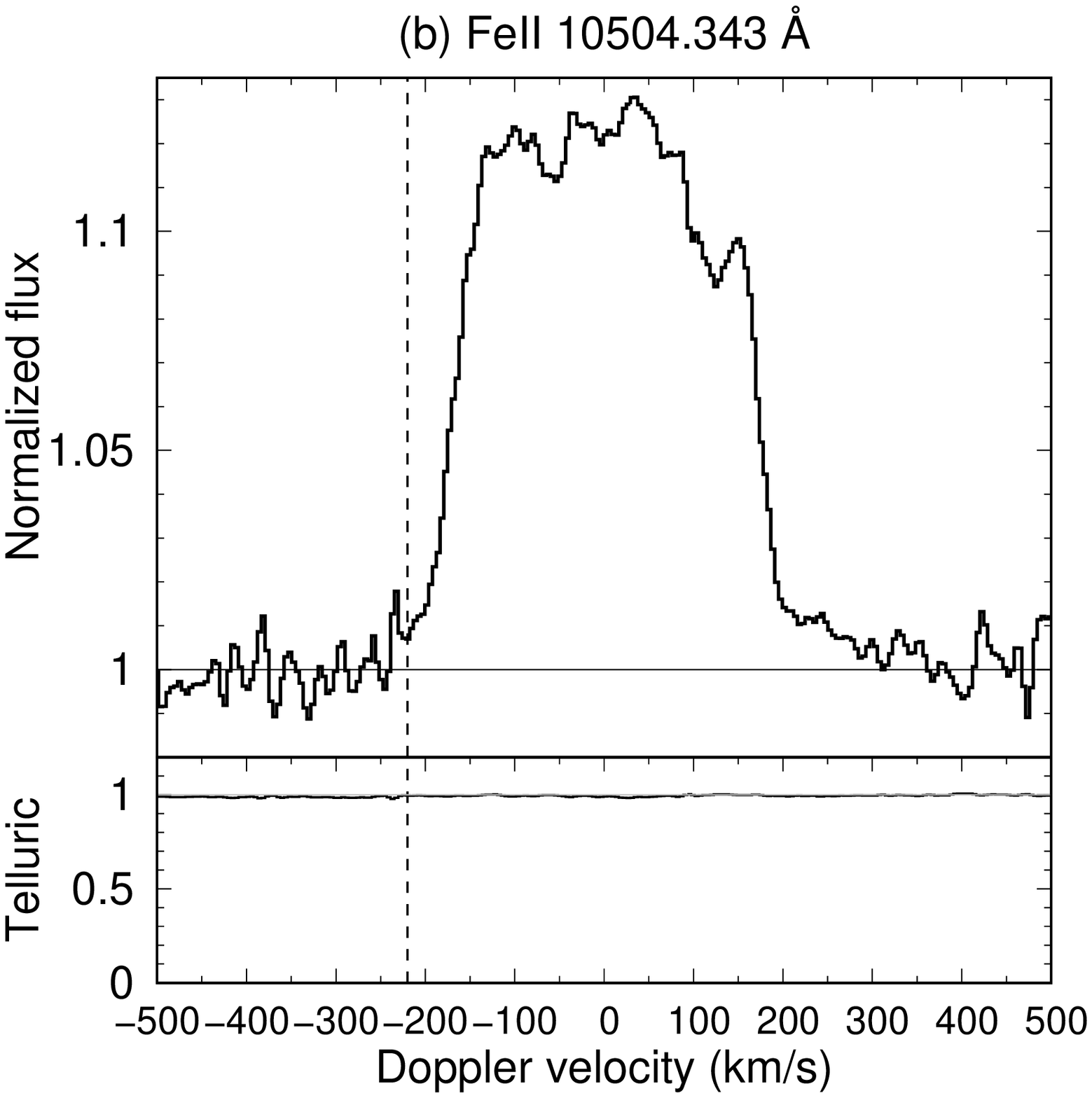}}
}
\subfigure{
\resizebox{7cm}{!}{\includegraphics[width=45mm,angle=270]{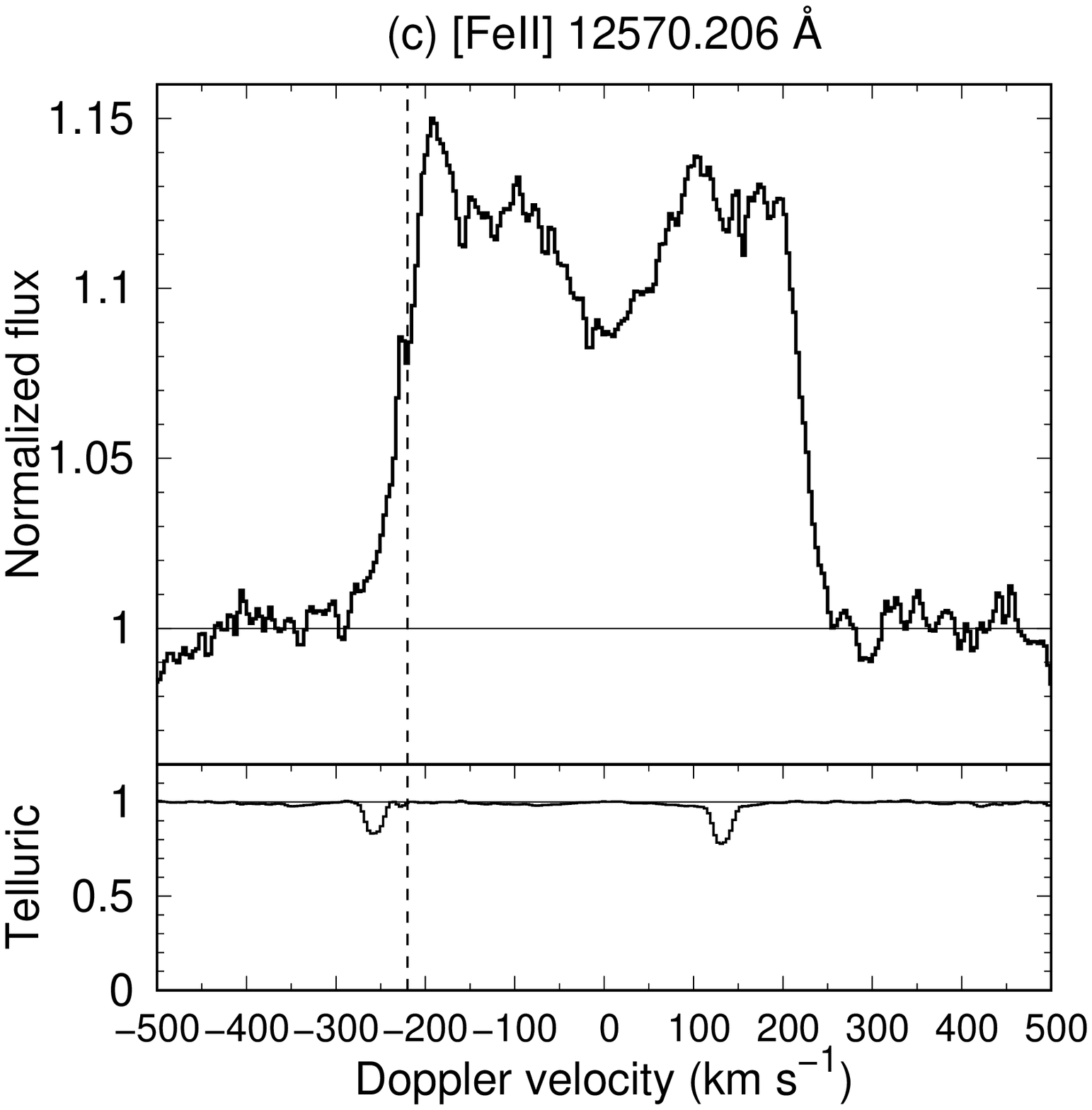}}
\resizebox{7cm}{!}{\includegraphics[width=45mm,angle=270]{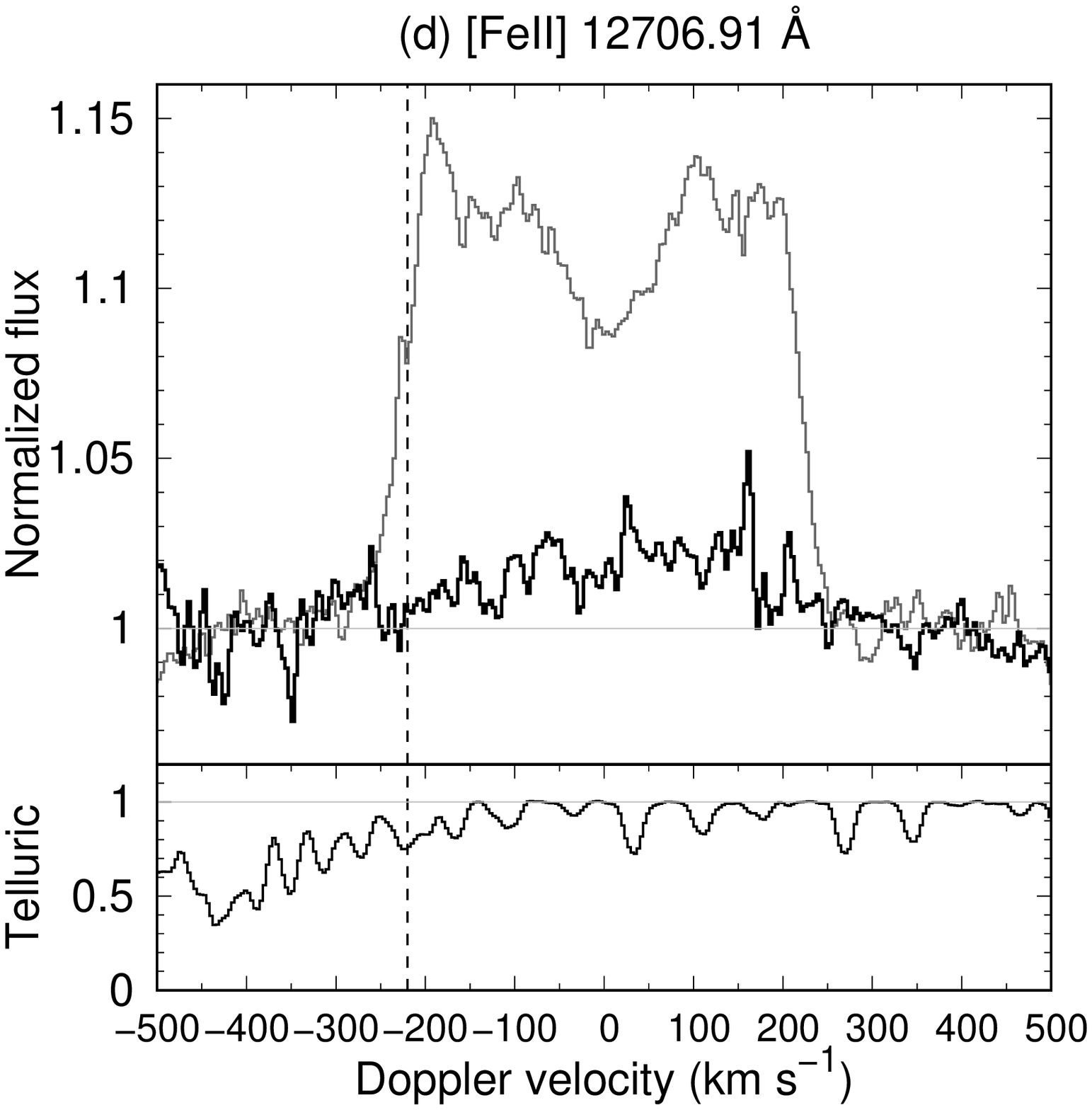}}
}
\subfigure{
\resizebox{7cm}{!}{\includegraphics[width=45mm,angle=270]{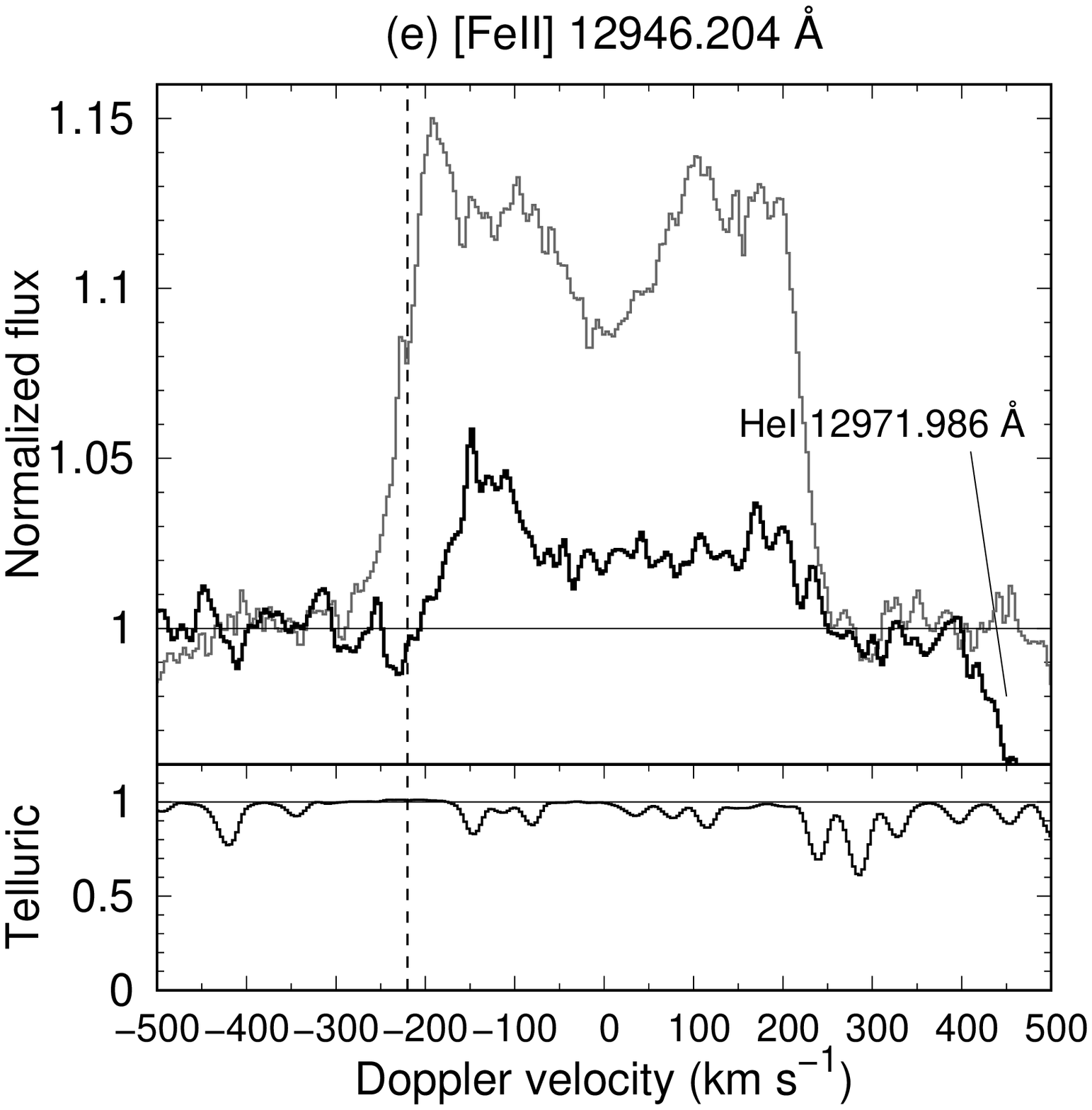}}
\resizebox{7cm}{!}{\includegraphics[width=45mm,angle=270]{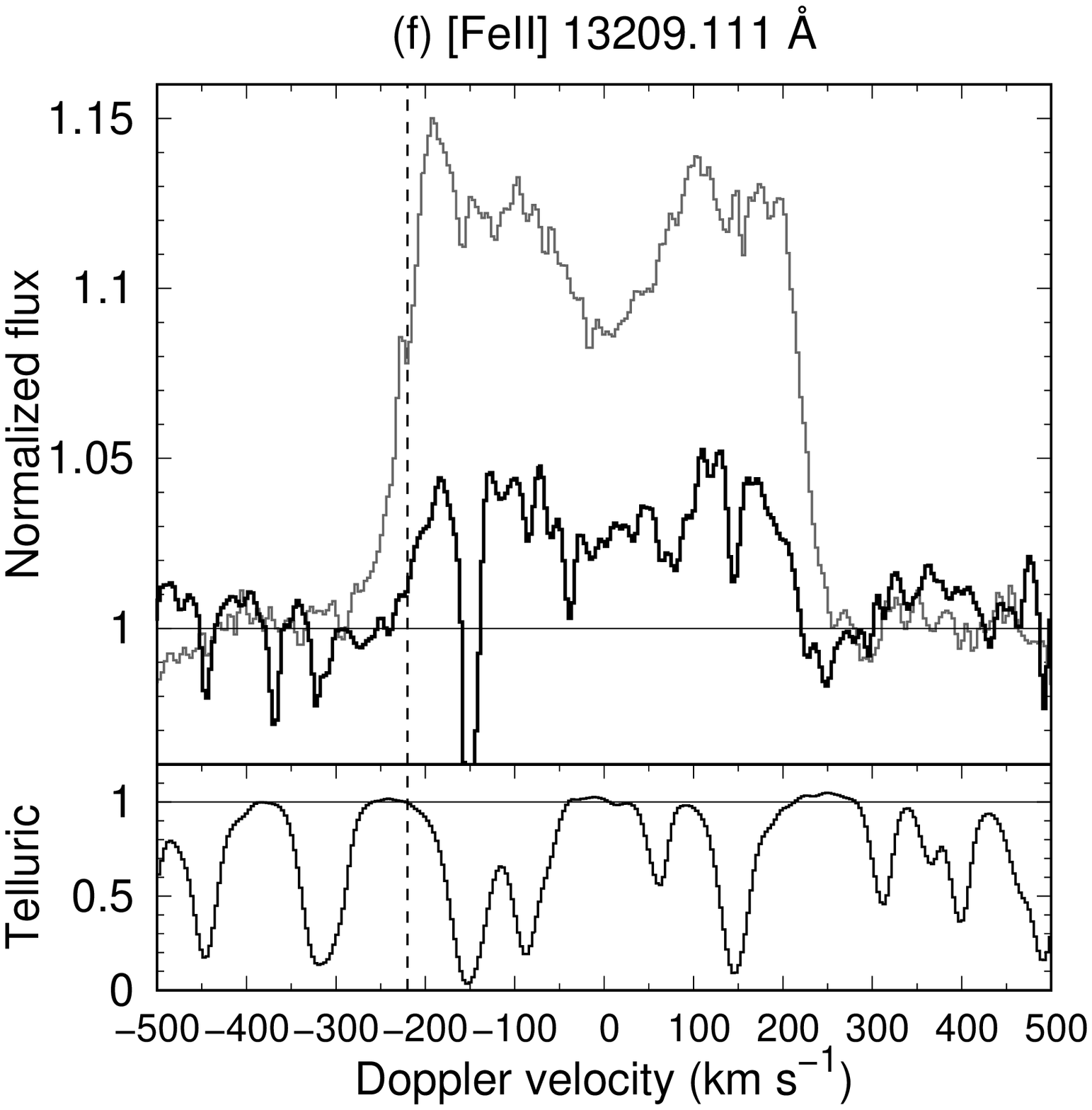}}
}
  \caption{Velocity profiles after the systemic velocity is corrected.
The profiles are normalized to fit the adjacent continuum to 1.
The spectra are telluric corrected.
The lower panels show profiles of the telluric standard star.
The grey profiles superposed on the panel (d)-(f) show that of [\ion{Fe}{II}] $\lambda$12570, adopted from the panel (c), which is the strongest among all the observed [\ion{Fe}{II}] lines.
The vertical dashed lines show $-220$~km~s$^{-1}$, which corresponds to the terminal velocity of the P Cygni profiles and the wind velocity of the [\ion{Fe}{II}] profiles.
}
\label{fig2}
  \end{center}
\end{figure*}

We found 90 emission lines in the whole spectrum of P Cygni.
Table \ref{tab:linelist} and Fig.\ \ref{spectrum} show the linelist and the whole spectrum we obtained.
We used VALD3 \citep{kup99} when identifying the atomic lines.
Fig.\ \ref{fig2} shows some characteristic velocity profiles of emission lines.
The so-called ``P Cygni'' profiles (Fig.\ \ref{fig2}a) are clearly seen in \ion{H}{I} and \ion{He}{I} lines with a terminal velocity of $-220$~km~s$^{-1}$,
which originates in the optically-thick and steady mass-loss wind region 
around the star with a radius of $\sim 100\,R_*$ (e.g., \citealt{lam85}; see also the inset in Fig.\ \ref{fig1}). 
Rounded profiles (Fig.\ \ref{fig2}b) with a velocity of $180$~km~s$^{-1}$ are visible in metal permitted lines, such as \ion{O}{I}, \ion{Mg}{II}, and \ion{Fe}{II},
which are formed in the optically-thin and steady mass-loss wind region with a radius of $\lesssim 100\,R_*$ (e.g., \citealt{naj97,mar97,ros01}).
The central velocity of \ion{Fe}{II} $\lambda10504$, $-28\pm1$~km~s$^{-1}$,
is consistent with the known systemic velocity of P Cygni (e.g., $-22$~km~s$^{-1}$ in \citealt{bar94} and \citealt{mar97}, $-27$~km~s$^{-1}$ in \citealt{mea01}, and $-29$~km~s$^{-1}$ in \citealt{sta93}).
The interpretation of ``P Cygni'' and ``rounded'' profiles is also consistent with those in previous studies.

We have clearly resolved for the first time each of the [\ion{Fe}{II}] emission lines in the obtained spectrum into two peaks, or a ``double-peak'' profile, with a velocity of $\simeq220$~km~s$^{-1}$ (Fig.\ \ref{fig2}c).
Several optical/infrared [\ion{Fe}{II}] lines were previously detected in this object, 
yet none of the lines have been resolved into multiple peaks, 
even though the observed profiles of the lines were known to be somewhat broad and probably to have a flat top at the (single) peak. 
\citet{mar00} proposed the interpretation that these ``flat-topped'' profiles are formed in the mass-loss wind region of a constant expansion velocity at a radius of $\sim100$ AU.
However, the compact emission region with a constant expansion velocity cannot generate such a ``double-peak'' profile as we obtained.
Figs.\ \ref{fig2}(c)--(f) show velocity profiles of all the four [\ion{Fe}{II}] lines obtained in this observation.
The [\ion{Fe}{II}] $\lambda$12570 ($a^4D_{7}-a^6D_{9}$) line is the strongest [\ion{Fe}{II}] line in the observed range of wavelengths, and clearly shows the ``double-peak'' profile at $v\simeq\pm220$~km~s$^{-1}$ and sub-peaks at $v\simeq\pm100$~km~s$^{-1}$.
The [\ion{Fe}{II}] $\lambda$12707 ($a^4D_{1}-a^6D_{1}$) line was detected, but too faint to evaluate the velocity profile quantitatively.
The [\ion{Fe}{II}] $\lambda$12946 ($a^4D_{5}-a^6D_{5}$) line was also detected, but its red wing is slightly contaminated with the adjacent \ion{He}{I} $\lambda$12972.
Although [\ion{Fe}{II}] $\lambda$13209 ($a^4D_{7}-a^6D_{7}$) is affected by residuals of a strong telluric absorption line,
the velocity profile shows the relatively clear ``double-peak'' profile in the same way as that of [\ion{Fe}{II}] $\lambda$12570.

\subsection{Spatial extent of [Fe II]} \label{sec:spatial}
\begin{figure}
\centering
\includegraphics[width=\columnwidth]{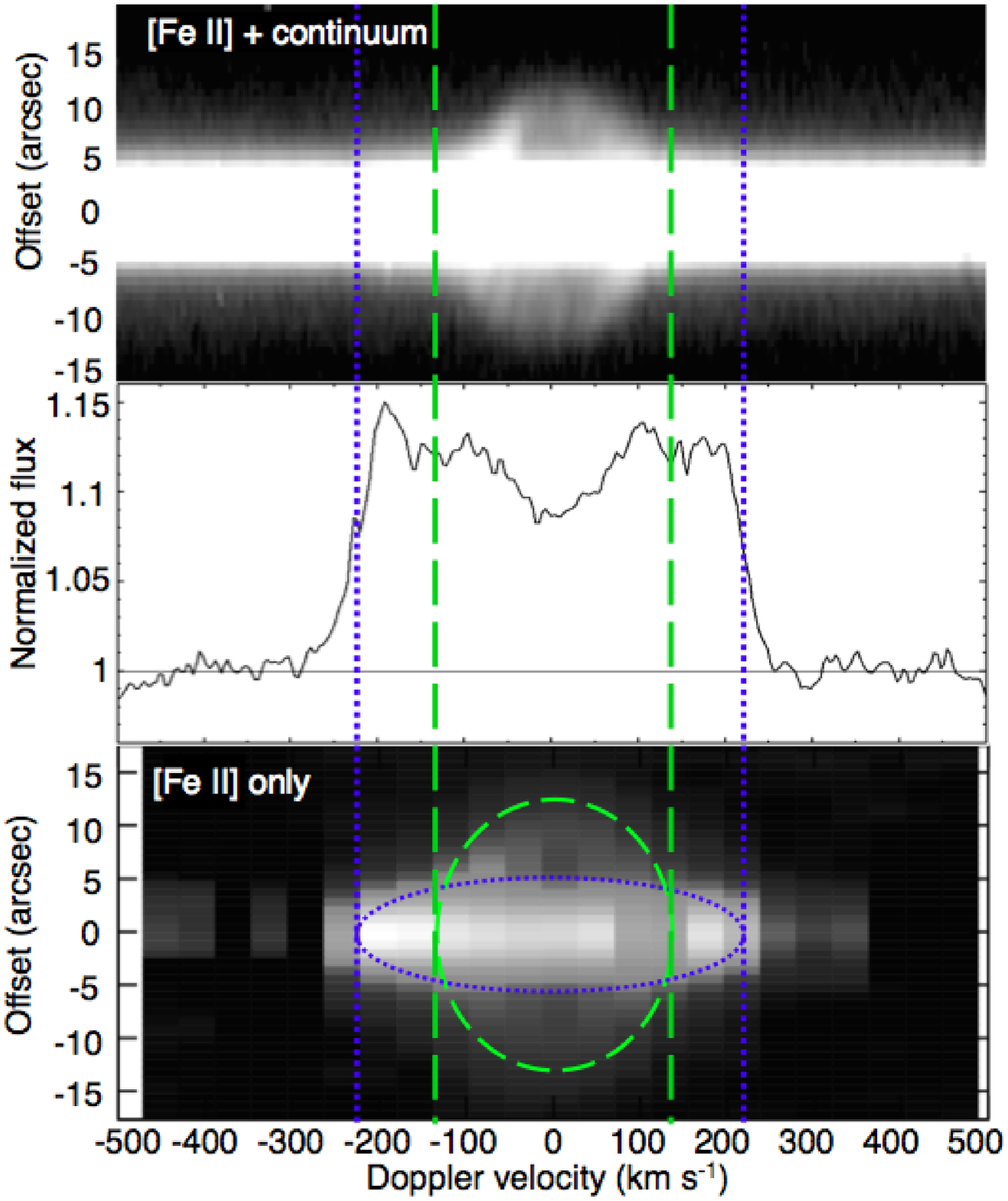} 
\caption{(Upper) Echelle image around the [\ion{Fe}{II}] $\lambda$12570 line. 
(Middle) Velocity profile of the [\ion{Fe}{II}] $\lambda$12570 line normalized to fit the adjacent continuum to 1, 
which was created using the data of $-5^{\prime\prime}$ to $+5^{\prime\prime}$ after telluric correction.
(Lower) Same image as the upper panel, but after the continuum subtraction.
Note that two telluric absorption lines at $v\simeq-270,+110$~km~s$^{-1}$ 
cause slightly weaker signal at the corresponding binned elements.
The blank areas at $v\simeq-370$~km~s$^{-1}$ and $v\gtrsim+360$~km~s$^{-1}$ have negative values after continuum subtractions due to statistical fluctuation.
The green dashed and blue dotted lines indicate $v=\pm140$~km~s$^{-1}$ and $v=\pm220$~km~s$^{-1}$, respectively.
}
\label{fig4}
\end{figure}

Fig.\ \ref{fig4} shows the echelle image around the [\ion{Fe}{II}] $\lambda$12570 line.
The 1600-AD outburst shell, which has an expansion velocity of $\simeq140$~km~s$^{-1}$ 
and a radius of $8^{\prime\prime}-10^{\prime\prime}$,
 is apparent in the echelle image
(n.b., this image is in the same configuration as in Fig.\ 5 of SH06 around [\ion{Fe}{II}] $\lambda$16440).
This shell is responsible to create the sub-peaks at $v\simeq\pm100$~km~s$^{-1}$ in the projected spectrum
(\S3.1, see also the green dotted lines in Fig.\ \ref{fig4}.)
However, the spatial extent of the stronger emission with double peak at $v\simeq\pm220$~km~s$^{-1}$ is unclear in the upper panel of Fig.\ \ref{fig4},
suggesting that it has a spatially smaller region than the outburst shell
and that it is hindered under the very strong continuum.

In order to study the spatial extent of the [\ion{Fe}{II}] emission component with $v\simeq\pm220$~km~s$^{-1}$ double peak,
we subtracted the model continuum on the echelle image.
First, we extracted 
52 one-dimensional spectra for each spatial pixel between $-15^{\prime\prime}$ and $+15^{\prime\prime}$
and fitted the continuum spectra on both sides of the [\ion{Fe}{II}] emission with a linear function.
We then subtracted the best-fit continuum spectra from the one-dimensional ones,
and reconstructed the [\ion{Fe}{II}] two-dimensional echelle image
from all of the 52 one-dimensional spectra.
In this reconstructed image,
we binned each set of 9 pixels ($=1.72$~\AA) along the wavelength axis 
in order to smear out stripe-like residuals on the continuum, 
which was generated as a side effect of pixelizing
the rotated spectra on the detector array.
As a result, 
the two velocity components are clearly separated into the two oval extensions in the spacio-velocity field, i.e.,
 the 1600-AD outburst shell component with $v\simeq140$~km~s$^{-1}$ (green dashed line in the figure) and 
the ``double-peak'' component with $v\simeq220$~km~s$^{-1}$ (blue dotted), in the lower panel of Fig.\ \ref{fig4}. 

\begin{figure}
\begin{center}
\includegraphics[angle=270,width=80mm]{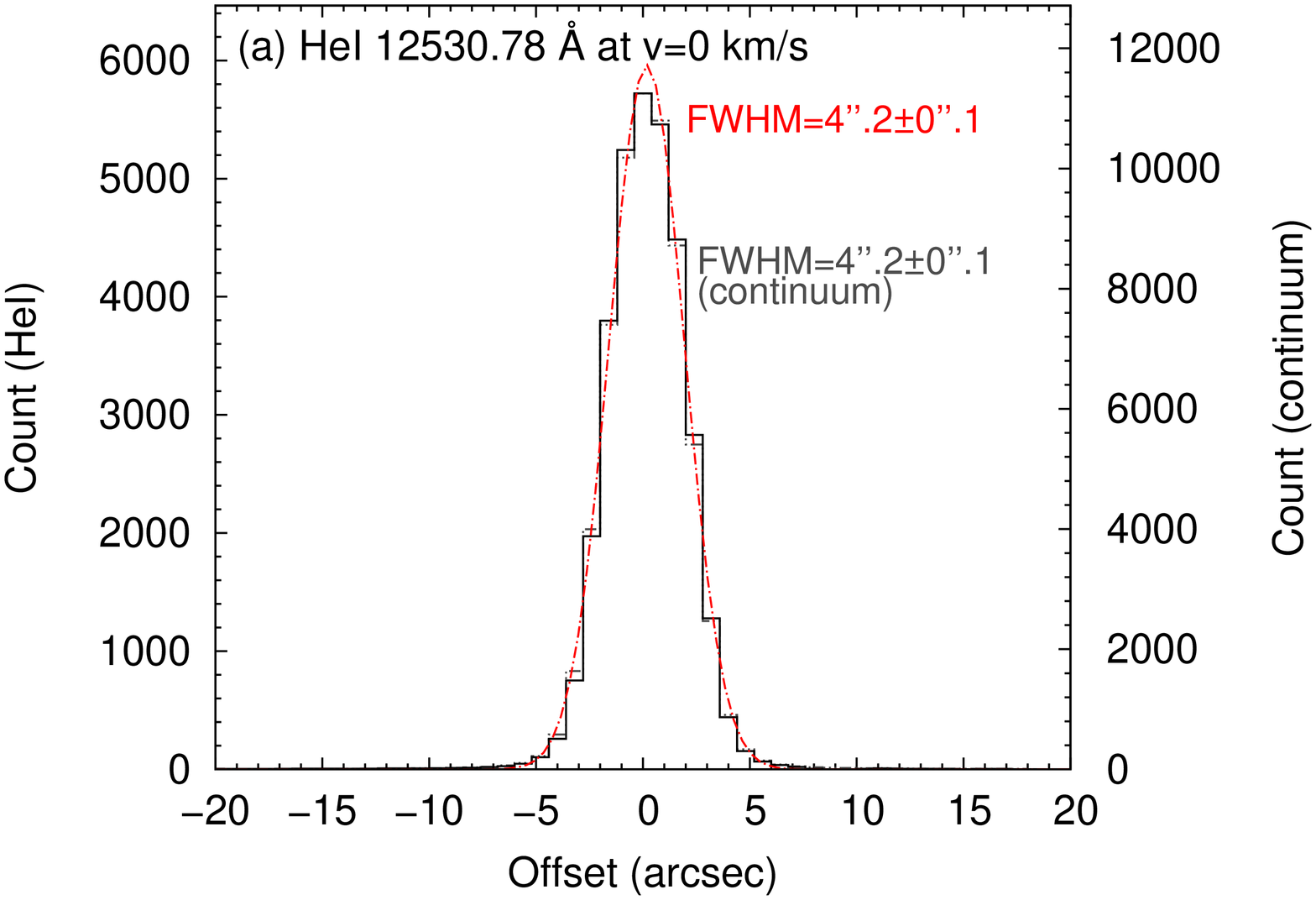}\\ 
\includegraphics[angle=270,width=80mm]{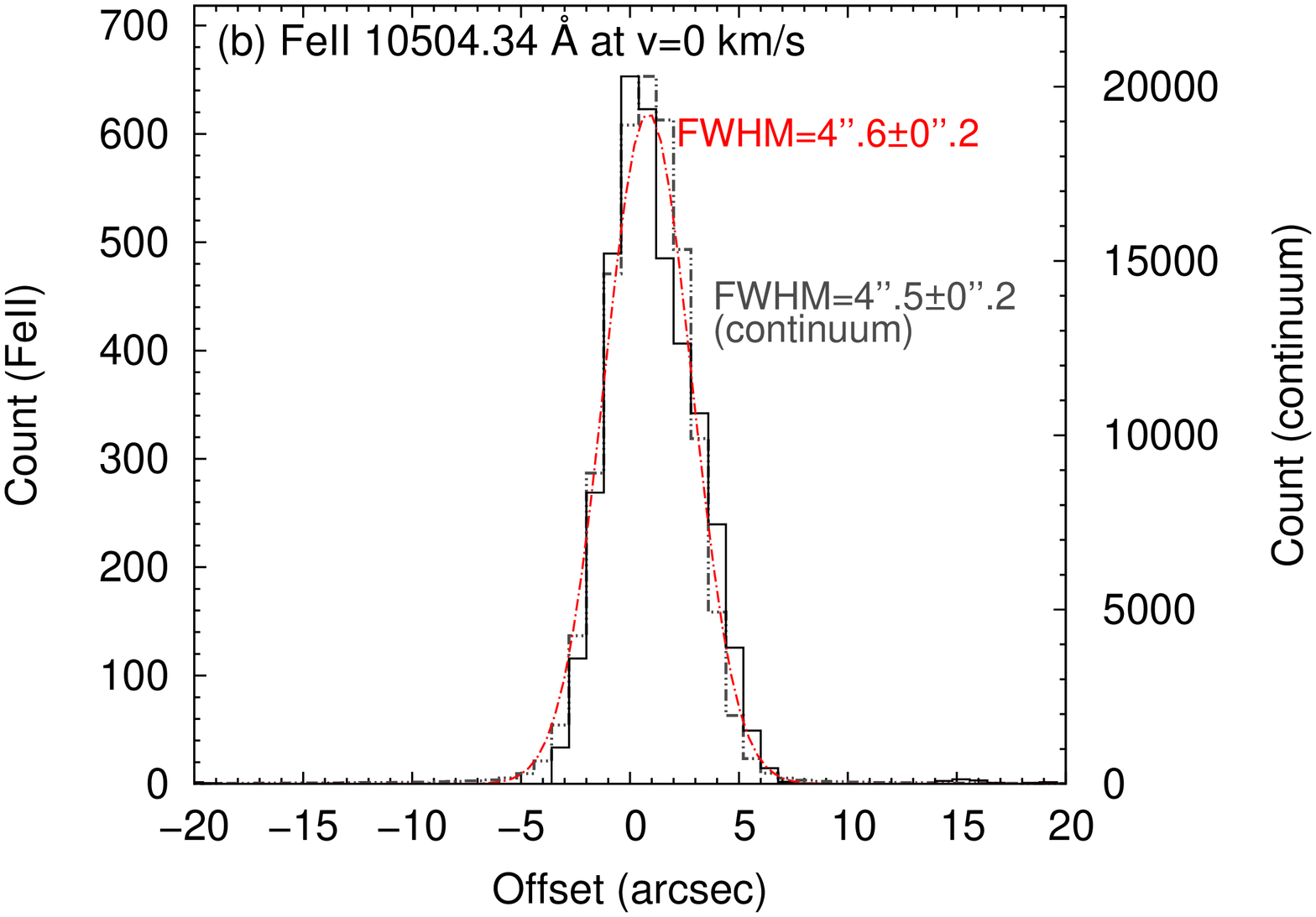} \\
\includegraphics[angle=270,width=80mm]{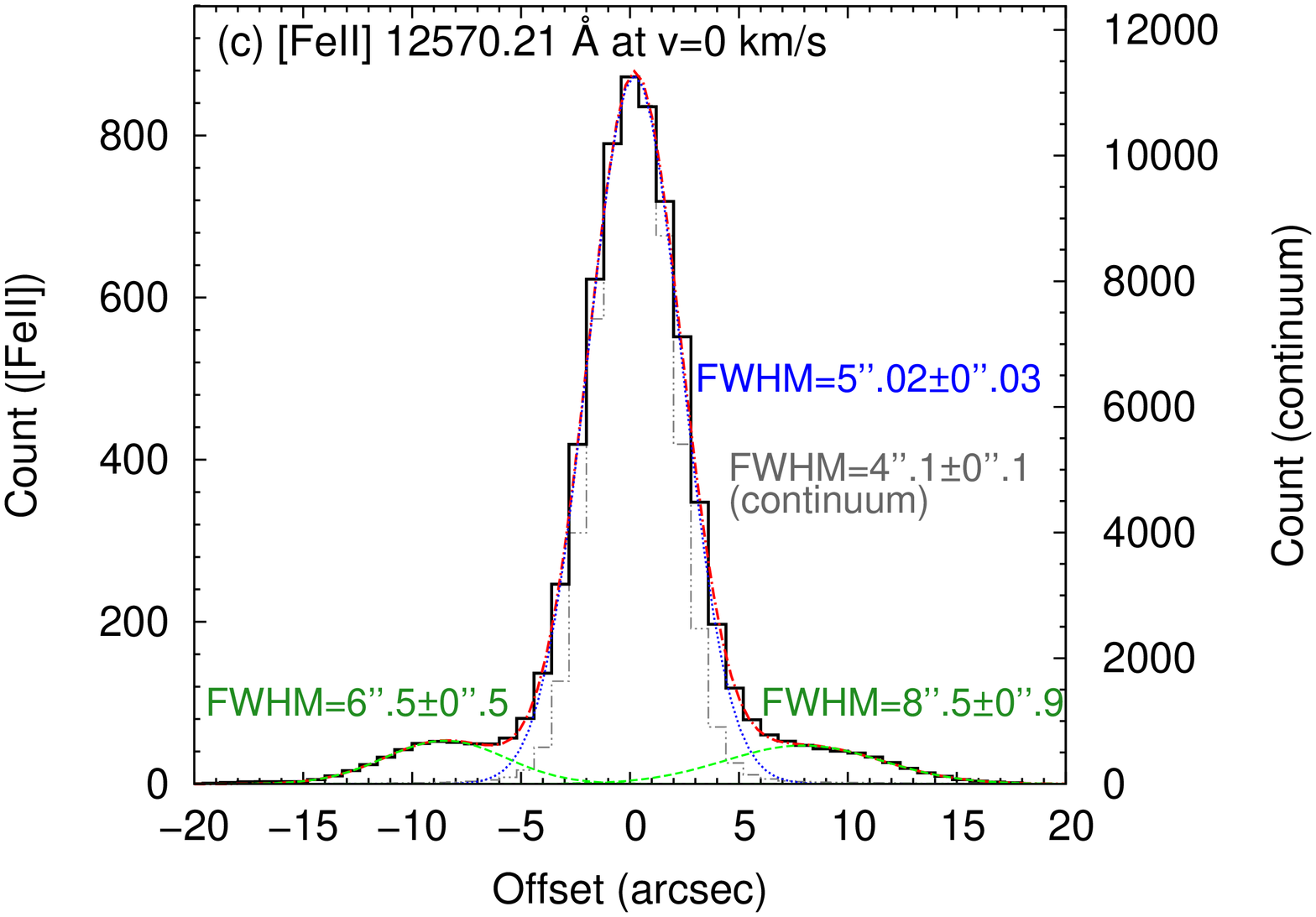} 
\end{center}
\caption{
Spatial profiles of three kinds of emission lines after continuum-subtraction: 
the \ion{He}{I} $\lambda$12530 line (a), \ion{Fe}{II} $\lambda$10504 line (b), and  [\ion{Fe}{II}] $\lambda$12570 line (c) at $v=0$~km~s$^{-1}$ (black solid).
The gray dot-dot-dashed lines show the continuum component,
whose scale in count is given on the right-hand axis in each panel.
In the upper panel, the green dashed lines show the Gaussian-fitted model for the 1600-AD outburst shell component,
the blue dotted line shows that for the ``double-peak'' component, and
the red dot-dashed line is the sum of the green and blue lines.
In the medium and lower panels, the red dot-dashed lines show the Gaussian-fitted model.
}
\label{fig6}
\end{figure}

We examined the spatial profiles of the {\it continuum-subtracted} echelle image
to quantitatively estimate the spatial extent of the emission regions.
The upper panel in Fig.\ \ref{fig6} shows the [\ion{Fe}{II}] spatial profile of the {\it continuum-subtracted} echelle image,
thus pure emission line echelle image, at $v=0$~km~s$^{-1}$.
We fitted the profile with three Gaussians:
One represents the ``double-peak'' component (blue dotted line in the figure) and the other two represent the blue- and redshifted components of the 1600-AD outburst shell (green dashed).
The spatial width of the ``double-peak'' component is 
$5^{\prime\prime}.02 \pm 0^{\prime\prime}.03$ at a full-width half-maximum (FWHM), 
which is clearly broader than that of the adjacent continuum component, $4^{\prime\prime}.1\pm0^{\prime\prime}.1$ (gray dot-dot-dashed),
which represents the seeing size because the continuum-emitting region is totally unresolved \citep{che00}.
This means that the [\ion{Fe}{II}] emission-line region is intrinsically spatially-extended.
By comparing the two spatial profiles,
the spatial extent even smaller than the seeing size can be examined (e.g., \citealt{tak09}).
Assuming that the extension of the emission region can be modeled with Gaussians,
the intrinsic FWHM of this component was estimated to be $\sqrt{(5^{\prime\prime}.02\pm0^{\prime\prime}.03)^2-(4^{\prime\prime}.1\pm0^{\prime\prime}.1)^2}=2^{\prime\prime}.9\pm0^{\prime\prime}.1$, 
which corresponds to $4.9\pm0.2\,\mathrm{kAU}$ or $(1.39\pm0.05)\times10^4\,R_*$ (see the dotted circle in Fig.\ \ref{fig1}).

For comparison, we made spatial profiles of \ion{He}{I} $\lambda$12531 (``P Cygni'' profile) and \ion{Fe}{II} $\lambda$10504 (``rounded'' profile) in the same way
from the {\it continuum-subtracted} echelle image at $v=0$~km~s$^{-1}$,
and plotted them in the medium and lower panels of Fig.\ \ref{fig6}.
The observed, adjacent, and intrinsic FWHM of \ion{He}{I} $\lambda$12531 are
$4^{\prime\prime}.2\pm0^{\prime\prime}.1$, $4^{\prime\prime}.2\pm0^{\prime\prime}.1$, and $0^{\prime\prime}\pm0^{\prime\prime}.6$, respectively, and
those of \ion{Fe}{II} $\lambda$10504 are $4^{\prime\prime}.6\pm0^{\prime\prime}.2$, $4^{\prime\prime}.5\pm0^{\prime\prime}.2$, and $1^{\prime\prime}.0\pm1^{\prime\prime}.3$, respectively.
This result shows that these lines are not extended spatially and are formed in a compact emission region close to the central star, as opposed to the [\ion{Fe}{II}] lines.

\begin{figure}
\centering
\includegraphics[angle=270,width=80mm]{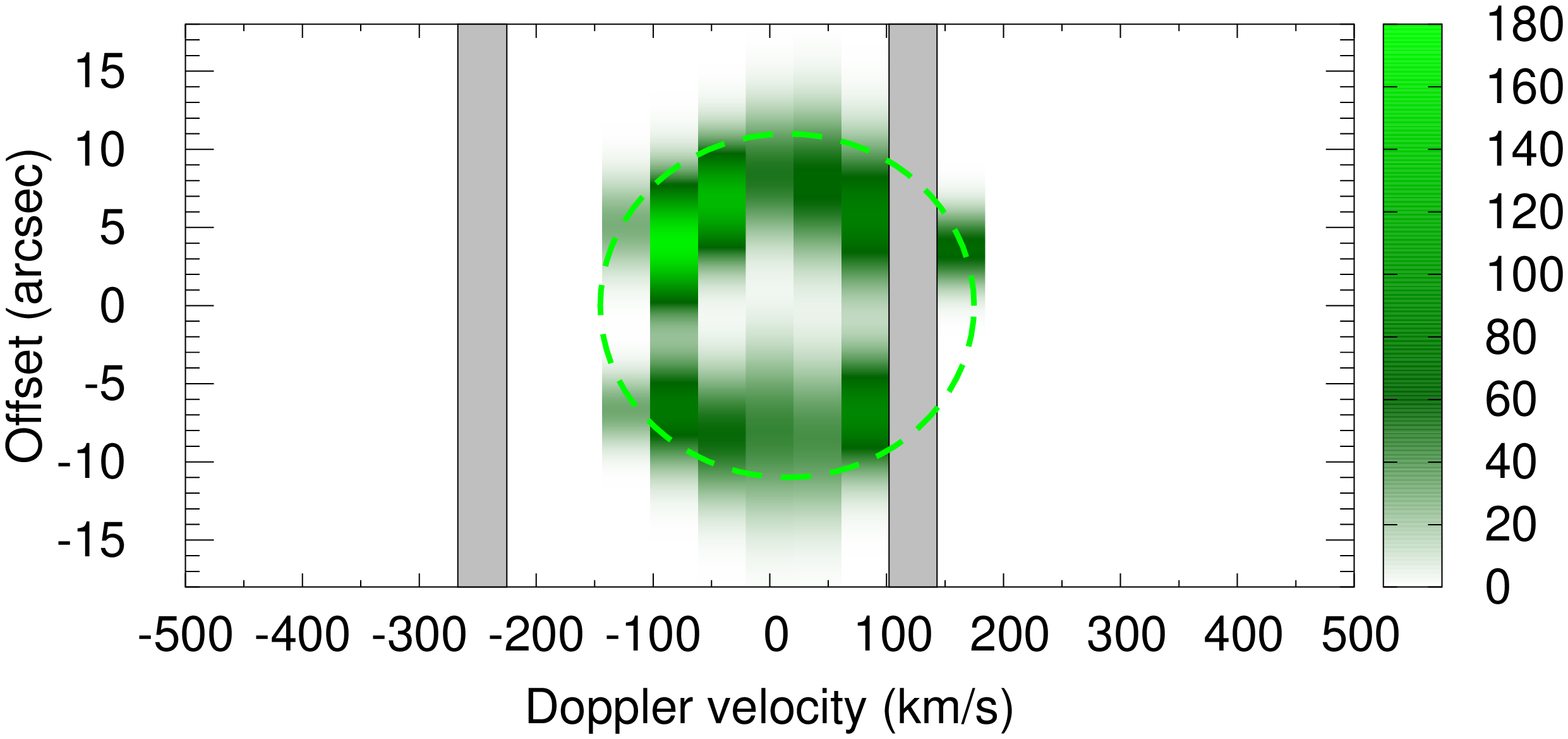} \\
\includegraphics[angle=270,width=80mm]{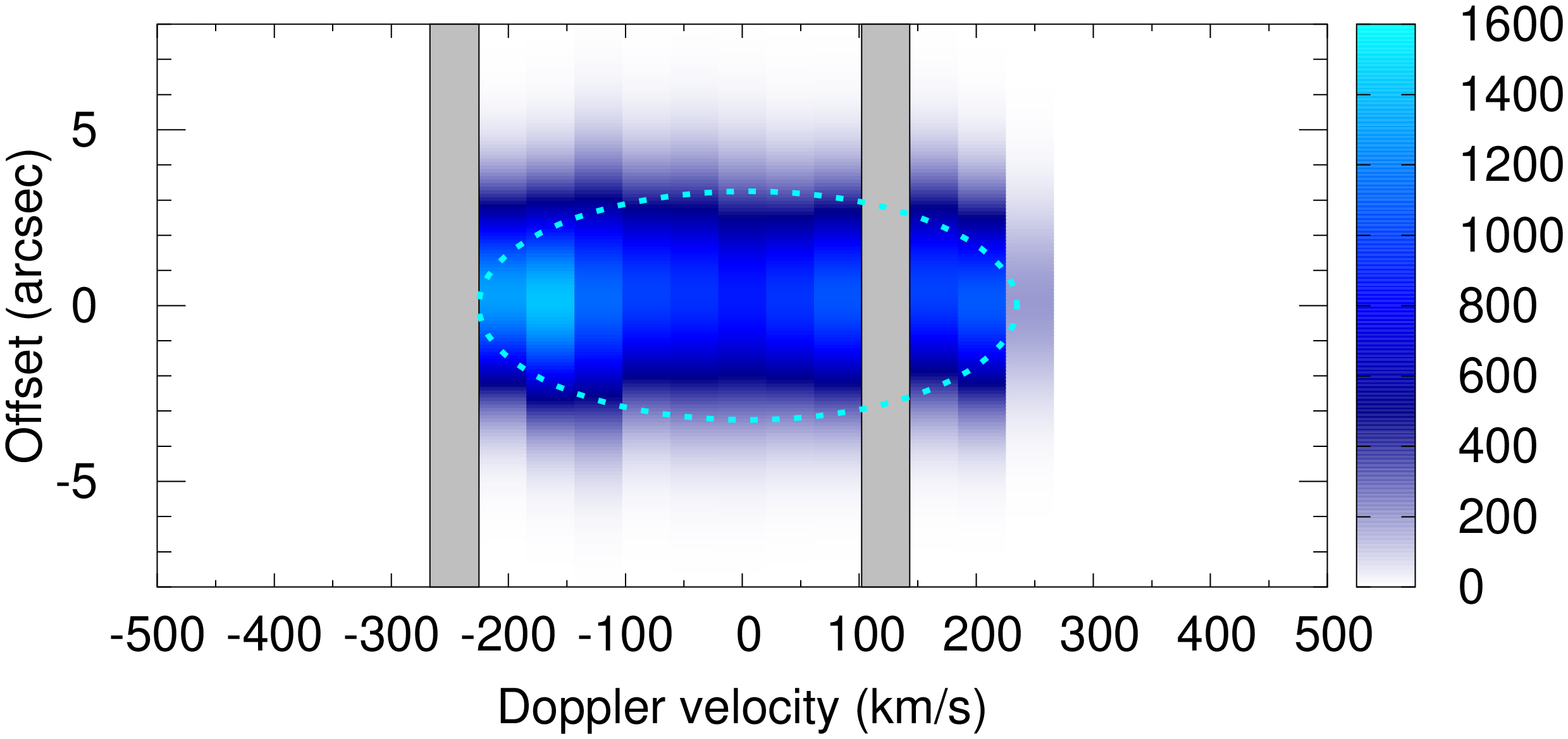} 
\caption{Reconstructed echelle images around the [\ion{Fe}{II}] $\lambda$12570 line. 
Green-tinted image (upper) shows the 1600-AD outburst shell component,
and blue-tinted image (lower) shows the extended emission region.
The telluric absorption regions at $v\simeq-270,+110$~km~s$^{-1}$ are not displayed.
The green-dashed and blue-dotted lines are same as Fig.\ \ref{fig4}.
}
\label{fig7}
\end{figure}

\begin{figure}
\centering
\includegraphics[angle=270,width=\columnwidth]{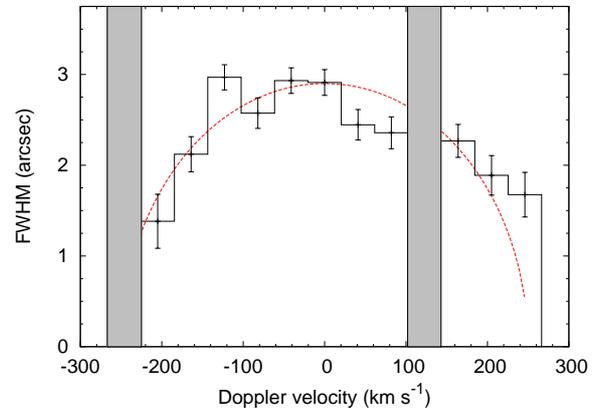} 
\caption{Intrinsic FWHM of the extended emission region.
The telluric absorption regions at $v\simeq-270,+110$~km~s$^{-1}$ are masked.
The red dashed line shows the model for spherical emission region.
}
\label{fig8}
\end{figure}

In order to unambiguously separate the ``double-peak'' component from those of the 1600-AD outburst shell,
we made spatial profiles of the echelle image for all the Doppler velocity bins from $-300$~km~s$^{-1}$ to $+300$~km~s$^{-1}$ with a 9 pixels ($=41$~km~s$^{-1}$) step,
and fitted all the components at $|v|\leq140$~km~s$^{-1}$ and at $|v|>140$~km~s$^{-1}$ with three and a single Gaussians, respectively (see Fig.\ \ref{fig6}c for the case of $|v|\leq140$~km~s$^{-1}$). 
The central Gaussian with the spatial offset of $0^{\prime\prime}$ indicates the ``double-peak'' component, 
whereas the two side Gaussians at $|v|\leq-140$~km~s$^{-1}$ indicate the 1600-AD outburst shell.
Next, we combined the fitted Gaussian profiles at all the Doppler velocity bins
in the spacio-velocity plane to reconstruct the echelle image.
Fig.\ \ref{fig7} shows the reconstructed images, which encompass
 the ``double-peak'' component (upper) and the 1600-AD outburst shell (lower).
Fig.\ \ref{fig8} shows the intrinsic FWHM of the ``double-peak'' component,
which is larger for smaller Doppler velocity.
The red dashed line in Fig.\ \ref{fig8} shows 
the model one when the emission region is spherical
($\mathrm{FWHM}=2^{\prime\prime}.9\sqrt{1-(v/220\,[\mathrm{km\,s^{-1}}])^2}$),
which is consistent with the trend of the observation.
Thus the emission region is found to be (quasi-)spherical.
Hereafter, we call the emission region of the ``double-peak'' component as the ``extended emission region'', in contrast to the 1600-AD outburst shell.

This ``extended emission region'' was, in fact, visible in SH06, though there was no mention about it in their paper.
The top panel of Fig.\ 4 in SH06 showed the intensity of [\ion{Fe}{II}] $\lambda$12567 and $\lambda$16435 lines\footnote{The observed wavelengths are used in SH06, whereas the vacuum wavelengths after the systemic velocity is corrected are used in this paper. Thus, the quoted wavelength values are slightly different between these two. For example, $\lambda12567$ in SH06 corresponds to $\lambda12570$ in this paper.}, where 
 the inner emission region appears in $-4^{\prime\prime}$ to $+4^{\prime\prime}$,
 whereas the 1600-AD outburst shell component appeared in $-10^{\prime\prime}$ to $-7^{\prime\prime}$ and $+8^{\prime\prime}$ to $+11^{\prime\prime}$.
We concluded that its inner emission region is identical to the ``extended emission region''
because the spatial extent of the regions is almost the same.
Consequently, we managed to constrain the upper limit of the radius of the [\ion{Fe}{II}] emission region, which has yet been unclear.


\section{Discussion}

\subsection{Origin of the [Fe II] ``double-peak'' profile}\label{sec:doublepeak}
When an emission line originates from an optically-thin spherically-symmetric shell/flow with a constant velocity,
the intensity per unit frequency bin is independent of frequency;
 as a result, the line shows a ``flat-topped'' profile \citep{app84,emerson,shu}.
The [\ion{Fe}{II}] velocity profiles of P Cygni have been treated as flat-topped in literature so far, and thus
the [\ion{Fe}{II}] emission region of P Cygni has been considered as a compact and spherically symmetric layer expanding with a constant velocity
(e.g., \citealt{sta91,isr91,mar97,mar00,kog07}).
However, our observation with high-resolution NIR spectroscopy revealed that the
[\ion{Fe}{II}] lines of P Cygni have ``double-peak'' profiles (\S\ref{result}).
Moreover, we found that the ``double-peak'' emission region was spatially extended by $4.9\pm0.2\,\mathrm{kAU}$ (\S\ref{sec:spatial}).

\begin{figure}
\centering
\includegraphics[width=\columnwidth,angle=270]{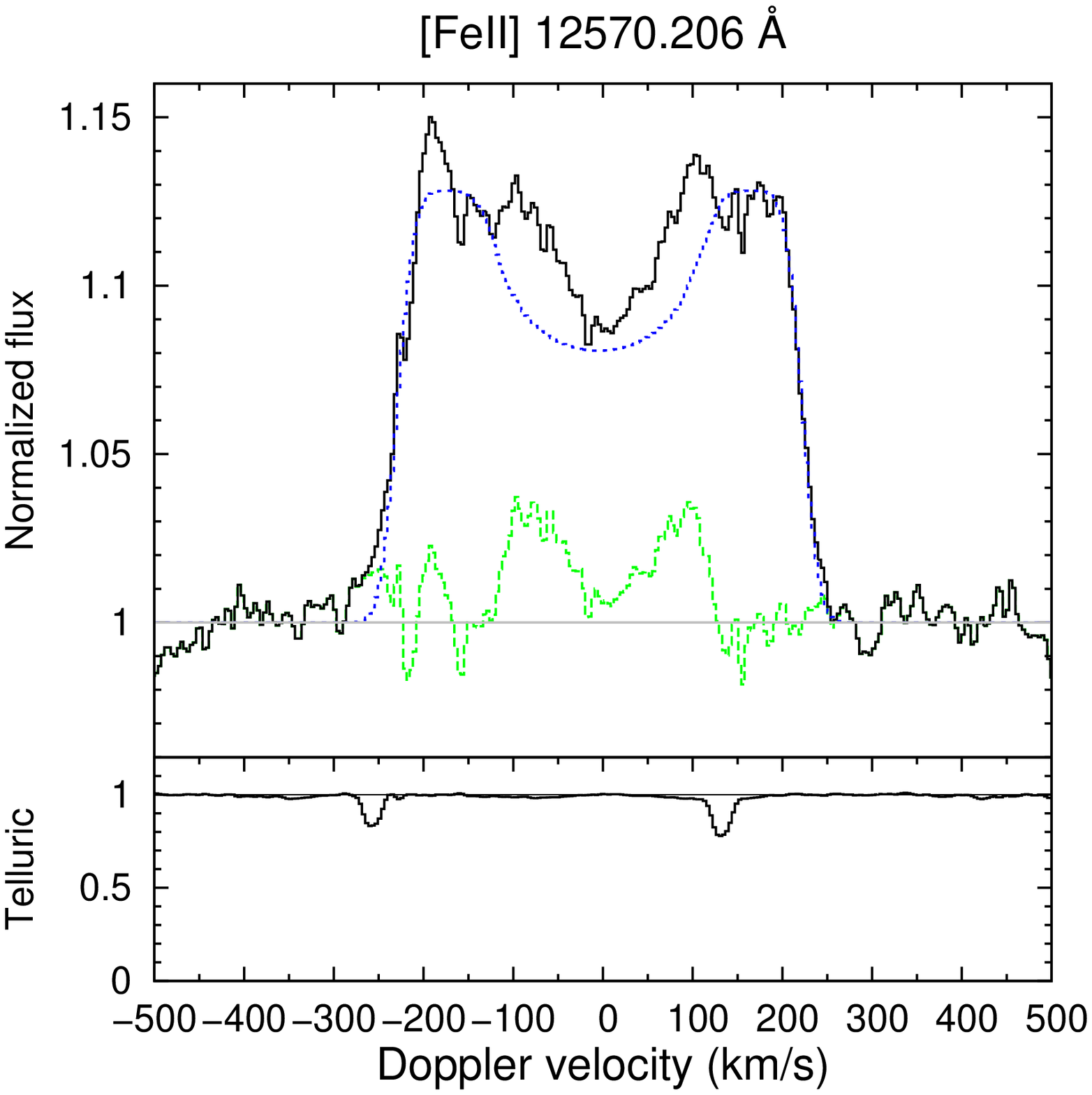} 
\caption{
The observed velocity profile of [\ion{Fe}{II}] $\lambda$12570 (black-solid; same as Fig.\ \ref{fig2}c) and the model one for the ``extended emission region'' (blue-dotted; see the main text for the detail of the model).
The green-dashed profile shows the residual component  of the model fitting.
}
\label{fig:fitting}
\end{figure}

Is it plausible for the ``double-peak'' profile to originate from the ``extended emission region''? 
In optical/infrared spectroscopy, when a part of an optically-thin spherically-symmetric shell/flow with a constant velocity is masked by a slit of spectrograph,
the intensity at the frequency corresponding to the radial velocity of the masked region becomes faint, and thus
a part of ``flat-topped'' velocity profile is scraped off (e.g., Chapter 9 of \citealt{shu}).
Therefore, when the emission region is extended wider than the slit width and
the extended component with a slow radial velocity is masked,
the velocity profile becomes hollow at the centre.
Assuming a spherical, homogeneous and isotropic emission region,
the model velocity profile can be calculated as a function of a diameter of the region.
The blue-dotted line in Fig.\ \ref{fig:fitting} shows the model velocity profile of
the emission region with a diameter of $2^{\prime\prime}$, which is slightly larger than the slit width ($1^{\prime\prime}.6$).
The expansion velocity of the model profile is set to be $220$~km~s$^{-1}$, and the instrumental resolution is fully taken into account.
Residuals seen in the velocity range of $-130$~km~s$^{-1}$ to $+130$~km~s$^{-1}$ (green-dashed) 
shows another clear double-peak profile, which is obviously a component of the 1600-AD outburst shell.

Indeed, the estimated diameter of the ``double-peak'' emission region (as defined with FWHM$=2^{\prime\prime}.9\pm0^{\prime\prime}.1$; see \S\ref{sec:spatial}) is similarly larger than the slit width ($1^{\prime\prime}.6$).
The emission region may be regarded as a region with uniform density with a diameter of roughly $2^{\prime\prime}$,
but more accurate modeling should await slit-scanning data for acquiring information on radial density distribution.
Consequently, the ``double-peak'' velocity profile is expected, as was observed, when the emission region is partially masked with the slit of spectrograph.
We must note that some intrinsic ununiformity of the emission region can enhance or skew the spectral features.

\subsection{Mass of the ``extended emission region''} \label{sec:mass}

Constraining the mass is essential to investigate the nature of the emission region.
Under the assumption that the ``extended emission region'' is spherically symmetric (\S\ref{sec:spatial}),
the total gas mass ($M$) is given by
\begin{equation}
M=\mu m_\mathrm{H} \frac{n_e}{f_\mathrm{H}}f\frac{4}{3}\pi(R_2^3-R_1^3),
\end{equation}
where $\mu$ is the mean molecular weight, $m_\mathrm{H}$ is the mass of hydrogen atom, $n_e$ is the number density of electron, $f_\mathrm{H}$ is the hydrogen ionization fraction, $f$ is the filling factor, and $R_2$ and $R_1$ are the outer and inner radii of the emission region, respectively (SH06).
The mean molecular weight $\mu$ is assumed to be 2.2.
$f_\mathrm{H}$ in the 1600-AD outburst shell is calculated to be 0.86, using the line ratio [\ion{N}{II}]/[\ion{N}{I}] in SH06.
In the ``extended emission region'', hydrogen atoms should be almost fully-ionized because they are much closer to the central star
than the 1600-AD outburst shell.
Therefore, we treated $f_\mathrm{H}$ in the ``extended emission region'' as unity.
The outer radius $R_2$ of the ``extended emission region'' is $1^{\prime\prime}.45$.
The inner radius $R_1$ must be small, given the spatial profile of the [\ion{Fe}{II}] $\lambda12570$ line (Fig.\ \ref{fig6}c) shows no apparent dip near the central star.
Here, we set $R_1=0$.
Estimating $f$ in the ``extended emission region'' is difficult because our observation could not spatially resolve its substructure in detail.
We assumed that $f$ is similar to that of the 1600-AD outburst shell in SH06, and adopt $f=0.2\pm0.1$.
$n_e$ can be calculated based on the ratios of the diagnostic [\ion{Fe}{II}] lines, $\lambda12567/\lambda16435$ and $\lambda15535/\lambda16435$ (SH06; see the second footnote\ for the notations).
In Fig.\ 4 of SH06, we read
$\lambda15535/\lambda16435$  as 0.16 in the ``extended emission region'' and as 0.1 in the 1600-AD outburst shell.
In SH06, $\lambda12567/\lambda16435$ was 1.3 throughout the regions, and 
thus $\lambda15535/\lambda12567$ is calculated to be 0.12 and 0.08 in the ``extended emission region'' and the 1600-AD outburst shell, respectively.
Comparing the $\lambda15535/\lambda12567$ ratios with the values in Fig.\ 3 of \citet{NS88}, 
we can estimate the respective electron densities to be $n_e=10000\,\mathrm{cm}^{-3}$ in the ``extended emission region'' and $6000\,\mathrm{cm}^{-3}$ in the 1600-AD outburst shell. 
The $n_e$ value hardly depends on the electron temperature in this  range.
With all the needed parameters (Table \ref{t1}), we derived the gas mass in the ``extended emission region'' to be $(8\pm4)\times10^{-4}M_\odot$.
This value is about 0.5\% of that in the 1600-AD outburst shell, $M=0.16\pm0.08\,M_\odot$.
The mass-loss rate with a constant stellar wind of P Cygni is $3\times10^{-5}M_\odot\,\mathrm{yr}^{-1}$ \citep{naj97b,naj97}; accordingly,
the ``extended emission region'' contains gas of the constant stellar wind within $30\pm15$ yr.
This value is consistent with a dynamical age, $R/\dot{R}=2.45\,\mathrm{kAU}/220$~km~s$^{-1}\sim50\,\mathrm{yr}$, which means that 
the total mass of this region can be explained by the steady stellar wind.
As a result, the ``extended emission region'' is not considered to be a remnant shell of an eruption but a stellar wind region.

\begin{table*}
 \begin{center}
 \caption{Parameters of the two [\ion{Fe}{II}] emission regions
 }\label{t1}
    \begin{threeparttable}
  \begin{tabular}{llcc}
   \hline
\hline
&& 1600-AD outburst shell\tnote{$\ast$1} & Extended emission region \\
\hline
Mean molecular weight &$\mu$ & \multicolumn{2}{c}{2.2} \\
Number density of electron&$n_e$ (cm$^{-3}$) & 6000 & 10000 \\
Hydrogen ionization fraction &$f_\mathrm{H}$ & 0.86 & 1 \\
Filling factor &$f$ & \multicolumn{2}{c}{$0.2\pm0.1$}  \\
Outer radius &$R_2$ (arcsec)& 9.7 & 1.45 \\
Inner radius &$R_1$ (arcsec)& 7.8 & 0 \\
\hline
Total gas mass &$M$ & $\simeq 0.16\pm0.08\,M_\odot$ & $\simeq (8\pm4)\times10^{-4}M_\odot$ \\
\hline
  \end{tabular}
\begin{tablenotes}\footnotesize
    \item[$\ast$1] The values in this column are adopted from SH06.
    \end{tablenotes}
  \end{threeparttable}
  \end{center}
\end{table*}

\subsection{[Fe II] excitation} \label{sec:nature}
The bright IR [\ion{Fe}{II}] emission is detected in all the observed LBVs with NIR spectroscopy \citep{smi02}.
For example, \citet{smi06} showed the [\ion{Fe}{II}] $\lambda$16435 spacio-velocity map of $\eta$ Carinae and
clearly indicated the two distinct emission regions within the nebulae;
one is the thicker skin extended inside of the H$_2$ shell and consistent with the ``Homunculus'', and the other is the inner region and consistent with the ``Little Homunculus'' \citep{ish03,smi05}.

The origin of the forbidden lines of LBV nebulae are not completely clear \citep{smi02,smi12}.
On one hand, the shocked stellar wind interacted with the ambient medium radiates the forbidden lines:
\citet{smi02} reported that the NIR [\ion{Fe}{II}] lines are good probes of shock-excited events such as LBV eruptions.
If the ``extended emission region'' of P Cygni is heated by shock,
its origin could be an episodic and relatively strong wind
that occasionally happens due to stellar fluctuations.
On the other hand, the forbidden lines may simply be radiatively excited \citep{smi06}:
\citet{smi07b} suggested that, compared with shock heating, radiative heating dominates the energy balance in the Homunculus of $\eta$ Carinae by 2 orders of magnitude.
If the ``extended emission region'' of P Cygni is radiatively heated,
what we see in the region is a simply steady stellar wind
with less number density and then is more extended than the compact stellar wind region traced by other lines.
As a result, both scenarios are consistent with the statement that
the ``extended emission region'' of P Cygni traces a stellar wind, not an eruption.
 
\subsection{Origin of the ``extended emission region''} 
The expansion velocity of the ``extended emission region'' ($\sim220$~km~s$^{-1}$) is consistent with the terminal velocity seen in the P Cygni profile (see the vertical lines in Fig.\ \ref{fig2}).
This implies that the ``extended emission region'' traces the outer wind after being accelerated and reaching the terminal velocity.
The expansion velocity of this region is faster than that of the 1600-AD outburst shell ($\sim140$~km~s$^{-1}$), suggesting that
this region overtakes the outer shell, produces a reverse shock, and emits the [\ion{Fe}{II}] emission.  
A numerical simulation in \citet{dwa98} shows that such a reverse shock emerges in the LBV nebulae (assuming $\eta$ Carinae), and its radius can be several times smaller than the outer shell, like this ``extended emission region''.
Consequently, we propose that the ``extended emission region'' traces the reverse shock region due to the stellar wind overtaking inside of the outburst shell.

Both the 1600-AD outburst shell region and the ``extended emission region'' show some brightness asymmetry in our data,
although we have discussed them with a spherically-symmetric geometry so far.
In the upper plot of Fig.\ \ref{fig7}, 
a bright emission was visible at the northern blueshifted part in the 1600-AD outburst shell\footnote{This feature is also seen in Fig.\ 5 of SH06 though the slit positions were not identical.}.
In the lower panel of Fig.\ \ref{fig7}, 
the blueshifted part of the ``extended emission region'' is found to be brighter than the other areas.
Because the 1600-AD outburst shell does not have velocity components at $v>140$~km~s$^{-1}$, 
this bright spot of the ``extended emission region'' is free from contamination of the shell.
As a result, the two regions may share the similar asymmetry although they are spatially separated.
This similarity also supports the reverse shock scenario;
the denser region in the 1600-AD shell naturally causes stronger emission in the shock as the wind overtakes the inside of the shell.

The extended [\ion{Fe}{II}] emission are also observed in the stellar wind region of $\eta$ Carinae.
\citet{smi04c} found a UV excess emission region at $0^{\prime\prime}.1-0^{\prime\prime}.6$ from the central star of $\eta$ Carinae, which emanates from the outer parts of the stellar wind region.
\citet{hil06} detected a UV [\ion{Fe}{II}] emission line from the UV region, $0^{\prime\prime}.2$ from the central star.
This [\ion{Fe}{II}] emission region in $\eta$ Carinae is similar to the ``extended emission region'' in P Cygni; both are spatially-extended around the central star and considered to be the outer parts of the stellar wind region. 
Moreover, their sizes are not so much different (0.2--1.4~kAU for $\eta$ Carinae and 4.9~kAU for P Cygni), and may scale with the luminosity of the central source.
Therefore, such extended [\ion{Fe}{II}] emission regions could be a common structure in LBV nebulae.

Structure of the P Cygni nebulae we propose in this paper is as follows.
The inner region of the wind ($\lesssim100\,R_*$) is traced by the metal permitted lines ($v=180$~km~s$^{-1}$).
The wind is accelerated by the radiation of the central star and reaches the terminal velocity at $\sim300\,R_*$ \citep{lam85}, which is traced by the P Cygni profile of H- and He emission lines ($v=220$~km~s$^{-1}$).
The wind velocity exceeds the expansion velocity of the 1600-AD outburst shell ($v=140$~km~s$^{-1}$), so that
the reverse shock emerges when the wind overtakes the shell, which produces the [\ion{Fe}{II}] ``extended emission region'' we have found in this paper ($v=220$~km~s$^{-1}$).
The radius and total gas mass of the shock region are about 15\% and 0.5\% of those of the 1600-AD outburst shell.
A promising way to study details of this newly-found region would be to spatially resolve it (within $3^{\prime\prime}$ from the central star) with adaptive-optics observations of the [\ion{Fe}{II}] lines.

\section*{acknowledgments}
We are grateful to the staff of Koyama Astronomical Observatory for their support during our observation. 
This study is financially supported by KAKENHI (16684001) Grant-in-Aid for Young Scientists (A), 
KAKENHI (20340042) Grant-in-Aid for Scientific Research (B), 
KAKENHI (26287028) Grant-in-Aid for Scientific Research (B), 
KAKENHI (21840052) Grant-in-Aid for Young Scientists (Start-up), 
and JSPS, MEXT - Supported program for the Strategic Research Foundation at Private Universities, 2008--2012, S0801061.
This work has made use of the VALD database, operated at Uppsala University, the Institute of Astronomy RAS in Moscow, and the University of Vienna.

\bibliographystyle{mn}
\bibliography{mn-jour,00}

\begin{thebibliography}{}
\makeatletter
\relax
\def\mn@urlcharsother{\let\do\@makeother \do\$\do\&\do\#\do\^\do\_\do\%\do\~}
\def\mn@doi{\begingroup\mn@urlcharsother \@ifnextchar [ {\mn@doi@}
  {\mn@doi@[]}}
\def\mn@doi@[#1]#2{\def\@tempa{#1}\ifx\@tempa\@empty \href
  {http://dx.doi.org/#2} {doi:#2}\else \href {http://dx.doi.org/#2} {#1}\fi
  \endgroup}
\def\mn@eprint#1#2{\mn@eprint@#1:#2::\@nil}
\def\mn@eprint@arXiv#1{\href {http://arxiv.org/abs/#1} {{\tt arXiv:#1}}}
\def\mn@eprint@dblp#1{\href {http://dblp.uni-trier.de/rec/bibtex/#1.xml}
  {dblp:#1}}
\def\mn@eprint@#1:#2:#3:#4\@nil{\def\@tempa {#1}\def\@tempb {#2}\def\@tempc
  {#3}\ifx \@tempc \@empty \let \@tempc \@tempb \let \@tempb \@tempa \fi \ifx
  \@tempb \@empty \def\@tempb {arXiv}\fi \@ifundefined
  {mn@eprint@\@tempb}{\@tempb:\@tempc}{\expandafter \expandafter \csname
  mn@eprint@\@tempb\endcsname \expandafter{\@tempc}}}

\bibitem[\protect\citeauthoryear{{Appenzeller}, {Oestreicher}  \&
  {Jankovics}}{{Appenzeller} et~al.}{1984}]{app84}
{Appenzeller} I.,  {Oestreicher} R.,   {Jankovics} I.,  1984, \aap, \href
  {http://adsabs.harvard.edu/abs/1984A%26A...141..108A} {141, 108}

\bibitem[\protect\citeauthoryear{{Arcidiacono} et~al.,}{{Arcidiacono}
  et~al.}{2014}]{arc14}
{Arcidiacono} C.,  et~al., 2014, \mn@doi [\mnras] {10.1093/mnras/stu1182},
  \href {http://adsabs.harvard.edu/abs/2014MNRAS.443.1142A} {443, 1142}

\bibitem[\protect\citeauthoryear{{Artigau}, {Martin}, {Humphreys}, {Davidson},
  {Chesneau}  \& {Smith}}{{Artigau} et~al.}{2011}]{art11}
{Artigau} {\'E}.,  {Martin} J.~C.,  {Humphreys} R.~M.,  {Davidson} K.,
  {Chesneau} O.,   {Smith} N.,  2011, \mn@doi [\aj]
  {10.1088/0004-6256/141/6/202}, \href
  {http://adsabs.harvard.edu/abs/2011AJ....141..202A} {141, 202}

\bibitem[\protect\citeauthoryear{{Barlow}, {Drew}, {Meaburn}  \&
  {Massey}}{{Barlow} et~al.}{1994}]{bar94}
{Barlow} M.~J.,  {Drew} J.~E.,  {Meaburn} J.,   {Massey} R.~M.,  1994, \mn@doi
  [\mnras] {10.1093/mnras/268.1.L29}, \href
  {http://adsabs.harvard.edu/abs/1994MNRAS.268L..29B} {268, L29}

\bibitem[\protect\citeauthoryear{{Chesneau} et~al.,}{{Chesneau}
  et~al.}{2000}]{che00}
{Chesneau} O.,  et~al., 2000, \mn@doi [\aaps] {10.1051/aas:2000227}, \href
  {http://adsabs.harvard.edu/abs/2000A%26AS..144..523C} {144, 523}

\bibitem[\protect\citeauthoryear{{Clark}, {Arkharov}, {Larionov}, {Ritchie},
  {Crowther}  \& {Najarro}}{{Clark} et~al.}{2011}]{cla11}
{Clark} J.~S.,  {Arkharov} A.,  {Larionov} V.,  {Ritchie} B.,  {Crowther} P.,
  {Najarro} F.,  2011, Bulletin de la Societe Royale des Sciences de Liege,
  \href {http://adsabs.harvard.edu/abs/2011BSRSL..80..361C} {80, 361}

\bibitem[\protect\citeauthoryear{{Dwarkadas} \& {Balick}}{{Dwarkadas} \&
  {Balick}}{1998}]{dwa98}
{Dwarkadas} V.~V.,  {Balick} B.,  1998, \mn@doi [\aj] {10.1086/300460}, \href
  {http://adsabs.harvard.edu/abs/1998AJ....116..829D} {116, 829}

\bibitem[\protect\citeauthoryear{{Emerson}}{{Emerson}}{1999}]{emerson}
{Emerson} D.,  1999, {Interpreting Astronomical Spectra}

\bibitem[\protect\citeauthoryear{{Groh}, {Damineli}  \& {Jablonski}}{{Groh}
  et~al.}{2007}]{gro07}
{Groh} J.~H.,  {Damineli} A.,   {Jablonski} F.,  2007, \mn@doi [\aap]
  {10.1051/0004-6361:20066401}, \href
  {http://adsabs.harvard.edu/abs/2007A%26A...465..993G} {465, 993}

\bibitem[\protect\citeauthoryear{{Gvaramadze} et~al.,}{{Gvaramadze}
  et~al.}{2012}]{gva12}
{Gvaramadze} V.~V.,  et~al., 2012, \mn@doi [\mnras]
  {10.1111/j.1365-2966.2012.20556.x}, \href
  {http://adsabs.harvard.edu/abs/2012MNRAS.421.3325G} {421, 3325}

\bibitem[\protect\citeauthoryear{{Hamann}, {Depoy}, {Johansson}  \&
  {Elias}}{{Hamann} et~al.}{1994}]{ham94}
{Hamann} F.,  {Depoy} D.~L.,  {Johansson} S.,   {Elias} J.,  1994, \mn@doi
  [\apj] {10.1086/173756}, \href
  {http://adsabs.harvard.edu/abs/1994ApJ...422..626H} {422, 626}

\bibitem[\protect\citeauthoryear{{Hartigan}, {Raymond}  \&
  {Pierson}}{{Hartigan} et~al.}{2004}]{har04}
{Hartigan} P.,  {Raymond} J.,   {Pierson} R.,  2004, \mn@doi [\apjl]
  {10.1086/425322}, \href {http://adsabs.harvard.edu/abs/2004ApJ...614L..69H}
  {614, L69}

\bibitem[\protect\citeauthoryear{{Hillier} et~al.,}{{Hillier}
  et~al.}{2006}]{hil06}
{Hillier} D.~J.,  et~al., 2006, \mn@doi [\apj] {10.1086/501225}, \href
  {http://adsabs.harvard.edu/abs/2006ApJ...642.1098H} {642, 1098}

\bibitem[\protect\citeauthoryear{{Humphreys}, {Davidson}  \&
  {Smith}}{{Humphreys} et~al.}{1999}]{hum99}
{Humphreys} R.~M.,  {Davidson} K.,   {Smith} N.,  1999, \mn@doi [\pasp]
  {10.1086/316420}, \href {http://adsabs.harvard.edu/abs/1999PASP..111.1124H}
  {111, 1124}

\bibitem[\protect\citeauthoryear{{Ikeda} et~al.,}{{Ikeda} et~al.}{2016}]{ike16}
{Ikeda} Y.,  et~al., 2016, in Ground-based and Airborne Instrumentation for
  Astronomy VI. p. 99085Z, \mn@doi{10.1117/12.2230886}

\bibitem[\protect\citeauthoryear{{Ishibashi} et~al.,}{{Ishibashi}
  et~al.}{2003}]{ish03}
{Ishibashi} K.,  et~al., 2003, \mn@doi [\aj] {10.1086/375306}, \href
  {http://adsabs.harvard.edu/abs/2003AJ....125.3222I} {125, 3222}

\bibitem[\protect\citeauthoryear{{Israelyan} \& {de Groot}}{{Israelyan} \& {de
  Groot}}{1991}]{isr91}
{Israelyan} G.~L.,  {de Groot} M.,  1991, \mn@doi [Astrophysics]
  {10.1007/BF01004784}, \href
  {http://adsabs.harvard.edu/abs/1991Ap.....34..171I} {34, 171}

\bibitem[\protect\citeauthoryear{{Kogure} \& {Leung}}{{Kogure} \&
  {Leung}}{2007}]{kog07}
{Kogure} T.,  {Leung} K.-C.,  eds, 2007, {The Astrophysics of Emission-Line
  Stars}  Astrophysics and Space Science Library Vol. 342

\bibitem[\protect\citeauthoryear{{Kupka}, {Piskunov}, {Ryabchikova}, {Stempels}
   \& {Weiss}}{{Kupka} et~al.}{1999}]{kup99}
{Kupka} F.,  {Piskunov} N.,  {Ryabchikova} T.~A.,  {Stempels} H.~C.,   {Weiss}
  W.~W.,  1999, \mn@doi [\aaps] {10.1051/aas:1999267}, \href
  {http://adsabs.harvard.edu/abs/1999A%26AS..138..119K} {138, 119}

\bibitem[\protect\citeauthoryear{{Lamers}}{{Lamers}}{1986}]{lam86}
{Lamers} H.~J.~G.~L.~M.,  1986, in {De Loore} C.~W.~H.,  {Willis} A.~J.,
  {Laskarides} P.,  eds,  IAU Symposium Vol. 116, Luminous Stars and
  Associations in Galaxies. pp 157--178

\bibitem[\protect\citeauthoryear{{Lamers}, {de Groot}  \&
  {Cassatella}}{{Lamers} et~al.}{1983}]{lam83}
{Lamers} H.~J.~G.~L.~M.,  {de Groot} M.,   {Cassatella} A.,  1983, \aap, \href
  {http://adsabs.harvard.edu/abs/1983A%26A...128..299L} {128, 299}

\bibitem[\protect\citeauthoryear{{Lamers}, {Korevaar}  \&
  {Cassatella}}{{Lamers} et~al.}{1985}]{lam85}
{Lamers} H.~J.~G.~L.~M.,  {Korevaar} P.,   {Cassatella} A.,  1985, \aap, \href
  {http://adsabs.harvard.edu/abs/1985A%26A...149...29L} {149, 29}

\bibitem[\protect\citeauthoryear{{Langer}, {Hamann}, {Lennon}, {Najarro},
  {Pauldrach}  \& {Puls}}{{Langer} et~al.}{1994}]{lan94}
{Langer} N.,  {Hamann} W.-R.,  {Lennon} M.,  {Najarro} F.,  {Pauldrach}
  A.~W.~A.,   {Puls} J.,  1994, \aap, \href
  {http://adsabs.harvard.edu/abs/1994A%26A...290..819L} {290}

\bibitem[\protect\citeauthoryear{{Lee}, {Moon}, {Koo}, {Lee}  \&
  {Matthews}}{{Lee} et~al.}{2009}]{lee09}
{Lee} H.-G.,  {Moon} D.-S.,  {Koo} B.-C.,  {Lee} J.-J.,   {Matthews} K.,  2009,
  \mn@doi [\apj] {10.1088/0004-637X/691/2/1042}, \href
  {http://adsabs.harvard.edu/abs/2009ApJ...691.1042L} {691, 1042}

\bibitem[\protect\citeauthoryear{{Markova}}{{Markova}}{2000}]{mar00}
{Markova} N.,  2000, \mn@doi [\aaps] {10.1051/aas:2000217}, \href
  {http://adsabs.harvard.edu/abs/2000A%26AS..144..391M} {144, 391}

\bibitem[\protect\citeauthoryear{{Markova} \& {de Groot}}{{Markova} \& {de
  Groot}}{1997}]{mar97}
{Markova} N.,  {de Groot} M.,  1997, \aap, \href
  {http://adsabs.harvard.edu/abs/1997A%26A...326.1111M} {326, 1111}

\bibitem[\protect\citeauthoryear{{Meaburn}}{{Meaburn}}{2001}]{mea01}
{Meaburn} J.,  2001, in {de Groot} M.,  {Sterken} C.,  eds,  Astronomical
  Society of the Pacific Conference Series Vol. 233, P Cygni 2000: 400 Years of
  Progress. p.~253

\bibitem[\protect\citeauthoryear{{Meaburn}, {Lopez}, {Barlow}  \&
  {Drew}}{{Meaburn} et~al.}{1996}]{mea96}
{Meaburn} J.,  {Lopez} J.~A.,  {Barlow} M.~J.,   {Drew} J.~E.,  1996, \mn@doi
  [\mnras] {10.1093/mnras/283.3.L69}, \href
  {http://adsabs.harvard.edu/abs/1996MNRAS.283L..69M} {283, L69}

\bibitem[\protect\citeauthoryear{{Meaburn}, {O'connor}, {L{\'o}pez}, {Bryce},
  {Redman}  \& {Noriega-Crespo}}{{Meaburn} et~al.}{2000}]{mea00}
{Meaburn} J.,  {O'connor} J.~A.,  {L{\'o}pez} J.~A.,  {Bryce} M.,  {Redman}
  M.~P.,   {Noriega-Crespo} A.,  2000, \mn@doi [\mnras]
  {10.1046/j.1365-8711.2000.03763.x}, \href
  {http://adsabs.harvard.edu/abs/2000MNRAS.318..561M} {318, 561}

\bibitem[\protect\citeauthoryear{{Meynet}, {Georgy}, {Hirschi}, {Maeder},
  {Massey}, {Przybilla}  \& {Nieva}}{{Meynet} et~al.}{2011}]{mey11}
{Meynet} G.,  {Georgy} C.,  {Hirschi} R.,  {Maeder} A.,  {Massey} P.,
  {Przybilla} N.,   {Nieva} M.-F.,  2011, Bulletin de la Societe Royale des
  Sciences de Liege, \href {http://adsabs.harvard.edu/abs/2011BSRSL..80..266M}
  {80, 266}

\bibitem[\protect\citeauthoryear{{Najarro}, {Kudritzki}, {Hillier}, {Lamers},
  {Voors}, {Morris}  \& {Waters}}{{Najarro} et~al.}{1997a}]{naj97b}
{Najarro} F.,  {Kudritzki} R.-P.,  {Hillier} D.~J.,  {Lamers} H.~J.~G.~L.~M.,
  {Voors} R.~H.~M.,  {Morris} P.~W.,   {Waters} L.~B.~F.~M.,  1997a, in {Nota}
  A.,  {Lamers} H.,  eds,  Astronomical Society of the Pacific Conference
  Series Vol. 120, Luminous Blue Variables: Massive Stars in Transition. p.~105

\bibitem[\protect\citeauthoryear{{Najarro}, {Hillier}  \& {Stahl}}{{Najarro}
  et~al.}{1997b}]{naj97}
{Najarro} F.,  {Hillier} D.~J.,   {Stahl} O.,  1997b, \aap, \href
  {http://adsabs.harvard.edu/abs/1997A%26A...326.1117N} {326, 1117}

\bibitem[\protect\citeauthoryear{{Nussbaumer} \& {Storey}}{{Nussbaumer} \&
  {Storey}}{1988}]{NS88}
{Nussbaumer} H.,  {Storey} P.~J.,  1988, \aap, \href
  {http://adsabs.harvard.edu/abs/1988A%26A...193..327N} {193, 327}

\bibitem[\protect\citeauthoryear{{Rossi}, {Muratorio}  \& {Viotti}}{{Rossi}
  et~al.}{2001}]{ros01}
{Rossi} C.,  {Muratorio} G.,   {Viotti} R.~F.,  2001, in {de Groot} M.,
  {Sterken} C.,  eds,  Astronomical Society of the Pacific Conference Series
  Vol. 233, P Cygni 2000: 400 Years of Progress. p.~105

\bibitem[\protect\citeauthoryear{{Shinn} et~al.,}{{Shinn} et~al.}{2013}]{shi13}
{Shinn} J.-H.,  et~al., 2013, \mn@doi [\apj] {10.1088/0004-637X/777/1/45},
  \href {http://adsabs.harvard.edu/abs/2013ApJ...777...45S} {777, 45}

\bibitem[\protect\citeauthoryear{{Shu}}{{Shu}}{1991}]{shu}
{Shu} F.~H.,  1991, {The physics of astrophysics. Volume 1: Radiation.}

\bibitem[\protect\citeauthoryear{{Smith}}{{Smith}}{2001}]{smi01a}
{Smith} N.,  2001, in {de Groot} M.,  {Sterken} C.,  eds,  Astronomical Society
  of the Pacific Conference Series Vol. 233, P Cygni 2000: 400 Years of
  Progress. p.~125

\bibitem[\protect\citeauthoryear{{Smith}}{{Smith}}{2002a}]{smi02}
{Smith} N.,  2002a, \mn@doi [\mnras] {10.1046/j.1365-8711.2002.05964.x}, \href
  {http://adsabs.harvard.edu/abs/2002MNRAS.336L..22S} {336, L22}

\bibitem[\protect\citeauthoryear{{Smith}}{{Smith}}{2002b}]{smi02c}
{Smith} N.,  2002b, \mn@doi [\mnras] {10.1046/j.1365-8711.2002.05966.x}, \href
  {http://adsabs.harvard.edu/abs/2002MNRAS.337.1252S} {337, 1252}

\bibitem[\protect\citeauthoryear{{Smith}}{{Smith}}{2005}]{smi05}
{Smith} N.,  2005, \mn@doi [\mnras] {10.1111/j.1365-2966.2005.08750.x}, \href
  {http://adsabs.harvard.edu/abs/2005MNRAS.357.1330S} {357, 1330}

\bibitem[\protect\citeauthoryear{{Smith}}{{Smith}}{2006}]{smi06}
{Smith} N.,  2006, \mn@doi [\apj] {10.1086/503766}, \href
  {http://adsabs.harvard.edu/abs/2006ApJ...644.1151S} {644, 1151}

\bibitem[\protect\citeauthoryear{{Smith}}{{Smith}}{2012}]{smi12}
{Smith} N.,  2012, in {Davidson} K.,  {Humphreys} R.~M.,  eds,  Astrophysics
  and Space Science Library Vol. 384, Eta Carinae and the Supernova Impostors.
  p.~145, \mn@doi{10.1007/978-1-4614-2275-4_7}

\bibitem[\protect\citeauthoryear{{Smith}}{{Smith}}{2014}]{smi14}
{Smith} N.,  2014, \mn@doi [\araa] {10.1146/annurev-astro-081913-040025}, \href
  {http://adsabs.harvard.edu/abs/2014ARA%26A..52..487S} {52, 487}

\bibitem[\protect\citeauthoryear{{Smith} \& {Ferland}}{{Smith} \&
  {Ferland}}{2007}]{smi07b}
{Smith} N.,  {Ferland} G.~J.,  2007, \mn@doi [\apj] {10.1086/510328}, \href
  {http://adsabs.harvard.edu/abs/2007ApJ...655..911S} {655, 911}

\bibitem[\protect\citeauthoryear{{Smith} \& {Hartigan}}{{Smith} \&
  {Hartigan}}{2006}]{smihar06}
{Smith} N.,  {Hartigan} P.,  2006, \mn@doi [\apj] {10.1086/498860}, \href
  {http://adsabs.harvard.edu/abs/2006ApJ...638.1045S} {638, 1045}

\bibitem[\protect\citeauthoryear{{Smith} et~al.,}{{Smith}
  et~al.}{2004}]{smi04c}
{Smith} N.,  et~al., 2004, \mn@doi [\apj] {10.1086/382185}, \href
  {http://adsabs.harvard.edu/abs/2004ApJ...605..405S} {605, 405}

\bibitem[\protect\citeauthoryear{{Smith}, {Li}, {Silverman}, {Ganeshalingam}
  \& {Filippenko}}{{Smith} et~al.}{2011}]{smi11}
{Smith} N.,  {Li} W.,  {Silverman} J.~M.,  {Ganeshalingam} M.,   {Filippenko}
  A.~V.,  2011, \mn@doi [\mnras] {10.1111/j.1365-2966.2011.18763.x}, \href
  {http://adsabs.harvard.edu/abs/2011MNRAS.415..773S} {415, 773}

\bibitem[\protect\citeauthoryear{{Stahl}, {Mandel}, {Szeifert}, {Wolf}  \&
  {Zhao}}{{Stahl} et~al.}{1991}]{sta91}
{Stahl} O.,  {Mandel} H.,  {Szeifert} T.,  {Wolf} B.,   {Zhao} F.,  1991, \aap,
  \href {http://adsabs.harvard.edu/abs/1991A%26A...244..467S} {244, 467}

\bibitem[\protect\citeauthoryear{{Stahl}, {Mandel}, {Wolf}, {Gaeng}, {Kaufer},
  {Kneer}, {Szeifert}  \& {Zhao}}{{Stahl} et~al.}{1993}]{sta93}
{Stahl} O.,  {Mandel} H.,  {Wolf} B.,  {Gaeng} T.,  {Kaufer} A.,  {Kneer} R.,
  {Szeifert} T.,   {Zhao} F.,  1993, \aaps, \href
  {http://adsabs.harvard.edu/abs/1993A%26AS...99..167S} {99, 167}

\bibitem[\protect\citeauthoryear{{Takami} et~al.,}{{Takami}
  et~al.}{2009}]{tak09}
{Takami} H.,  et~al., 2009, \mn@doi [\pasj] {10.1093/pasj/61.4.623}, \href
  {http://adsabs.harvard.edu/abs/2009PASJ...61..623T} {61, 623}

\bibitem[\protect\citeauthoryear{{Weis}}{{Weis}}{2011}]{wei11}
{Weis} K.,  2011, Bulletin de la Societe Royale des Sciences de Liege, \href
  {http://adsabs.harvard.edu/abs/2011BSRSL..80..440W} {80, 440}

\bibitem[\protect\citeauthoryear{{Yoshikawa}, {Ikeda}, {Fujishiro}, {Ichizawa},
  {Arai}, {Isogai}, {Yonehara}  \& {Kawakita}}{{Yoshikawa}
  et~al.}{2012}]{yos12}
{Yoshikawa} T.,  {Ikeda} Y.,  {Fujishiro} N.,  {Ichizawa} S.,  {Arai} A.,
  {Isogai} M.,  {Yonehara} A.,   {Kawakita} H.,  2012, in Ground-based and
  Airborne Telescopes IV. p. 84446G, \mn@doi{10.1117/12.925428}

\makeatother
\end{thebibliography}

\newpage
\begin{figure*}
\centering
\includegraphics[angle=270,width=160mm]{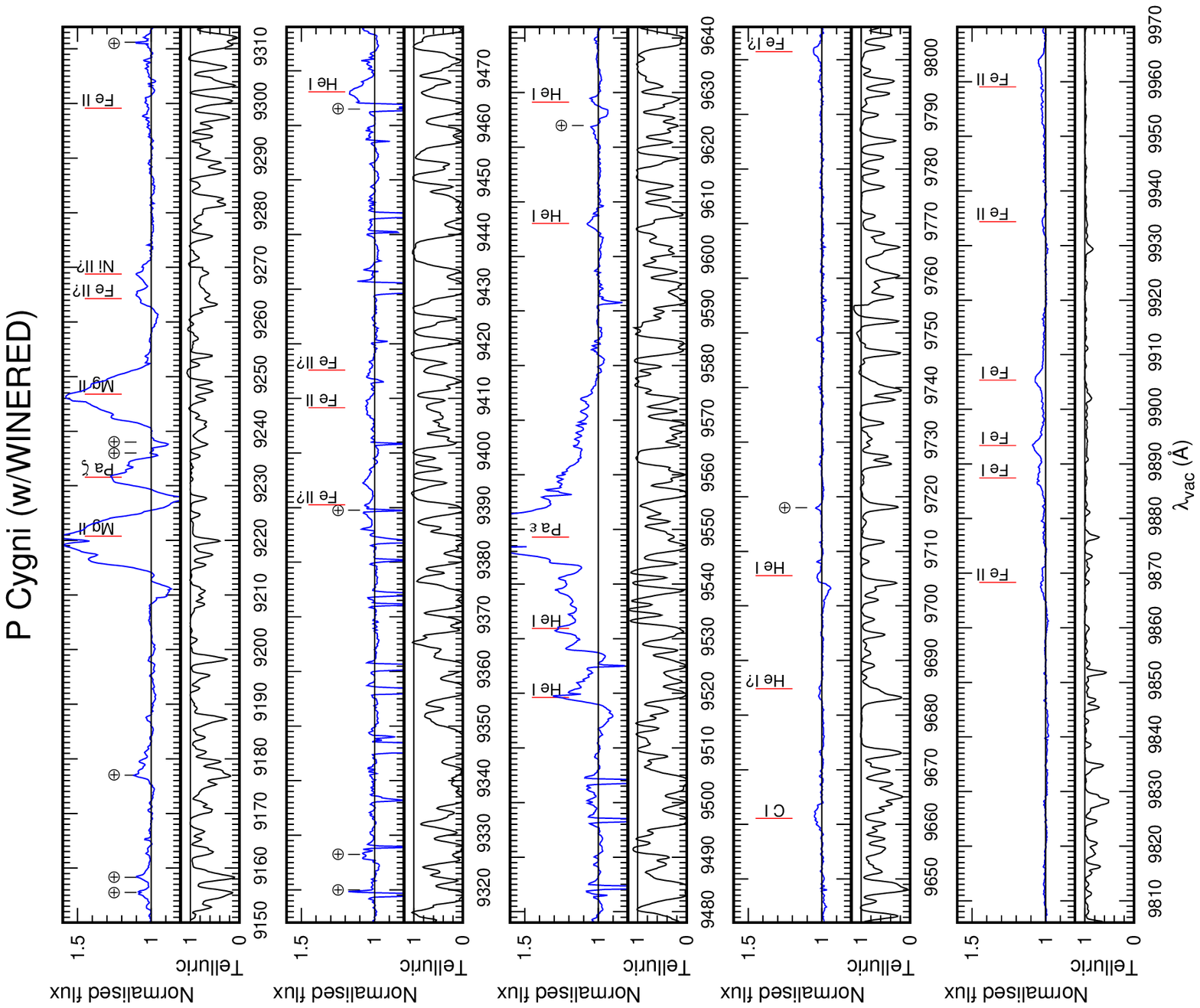}  
\caption{
The WINERED spectrum of P Cygni and its nebula in the slit area shown in Fig.\ \ref{fig1}. 
The horizontal axis shows the wavelength in vacuum. 
The telluric absorption is corrected using the spectrum of the standard star, 
which is shown in the bottom of each plot after removing the intrinsic stellar absorption lines and being normalized. 
The vertical red lines above the spectrum show emission lines.
The other spike-like emission features 
as well as a group of sharp absorption features in the bad atmospheric transmission region (especially 9310--9800\AA\ and 11120--11750\AA)
are due to incomplete telluric correction.
The $\oplus$ symbols show the part where telluric absorptions are not fully corrected because of their heavy absorption.
}
\label{spectrum}
\end{figure*}

\addtocounter{figure}{-1}
\begin{figure*}
\centering
\includegraphics[angle=270,width=160mm]{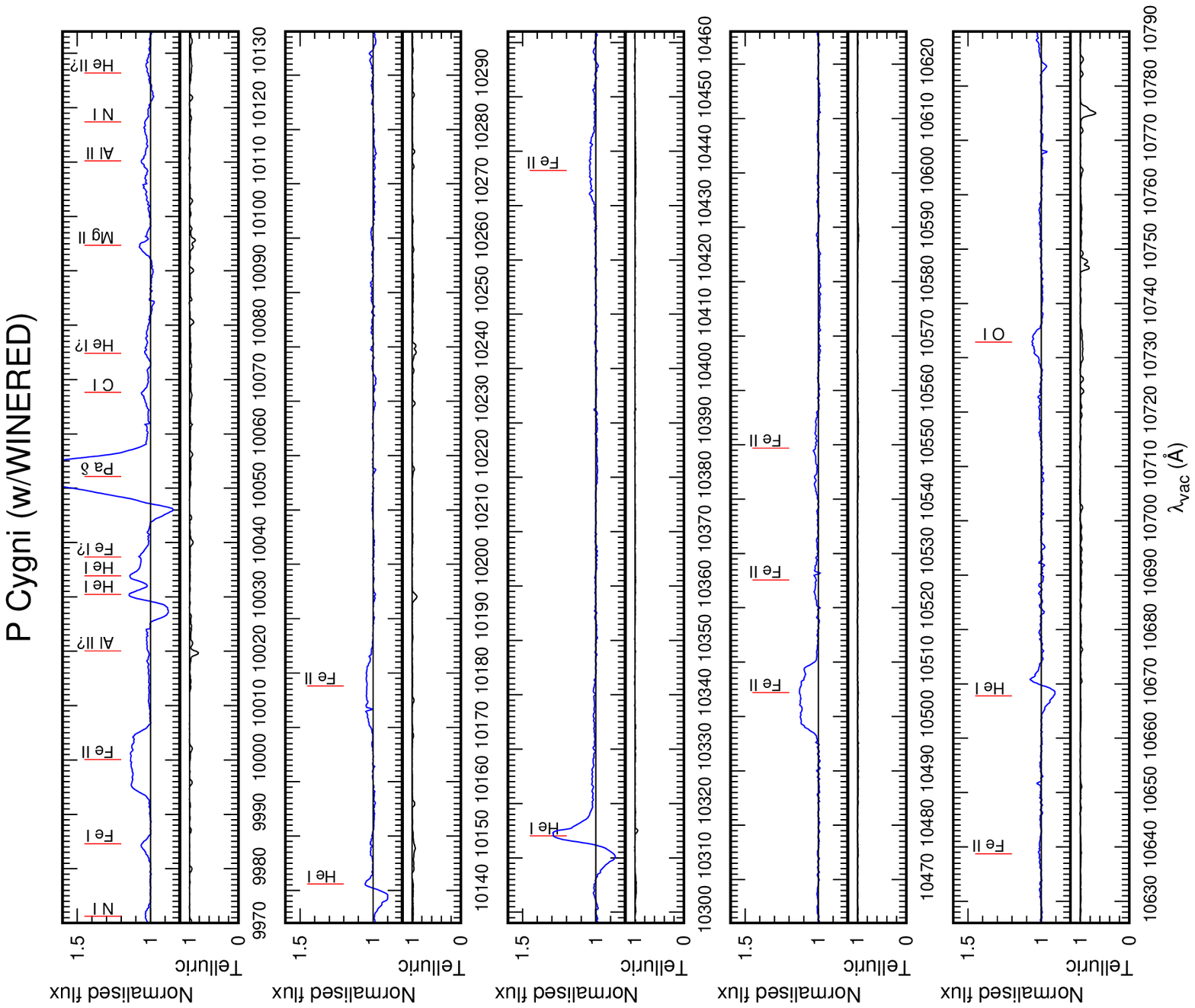}  
\caption{
{\it continued}
}
\label{spectrum2}
\end{figure*}

\addtocounter{figure}{-1}
\begin{figure*}
\centering
\includegraphics[angle=270,width=160mm]{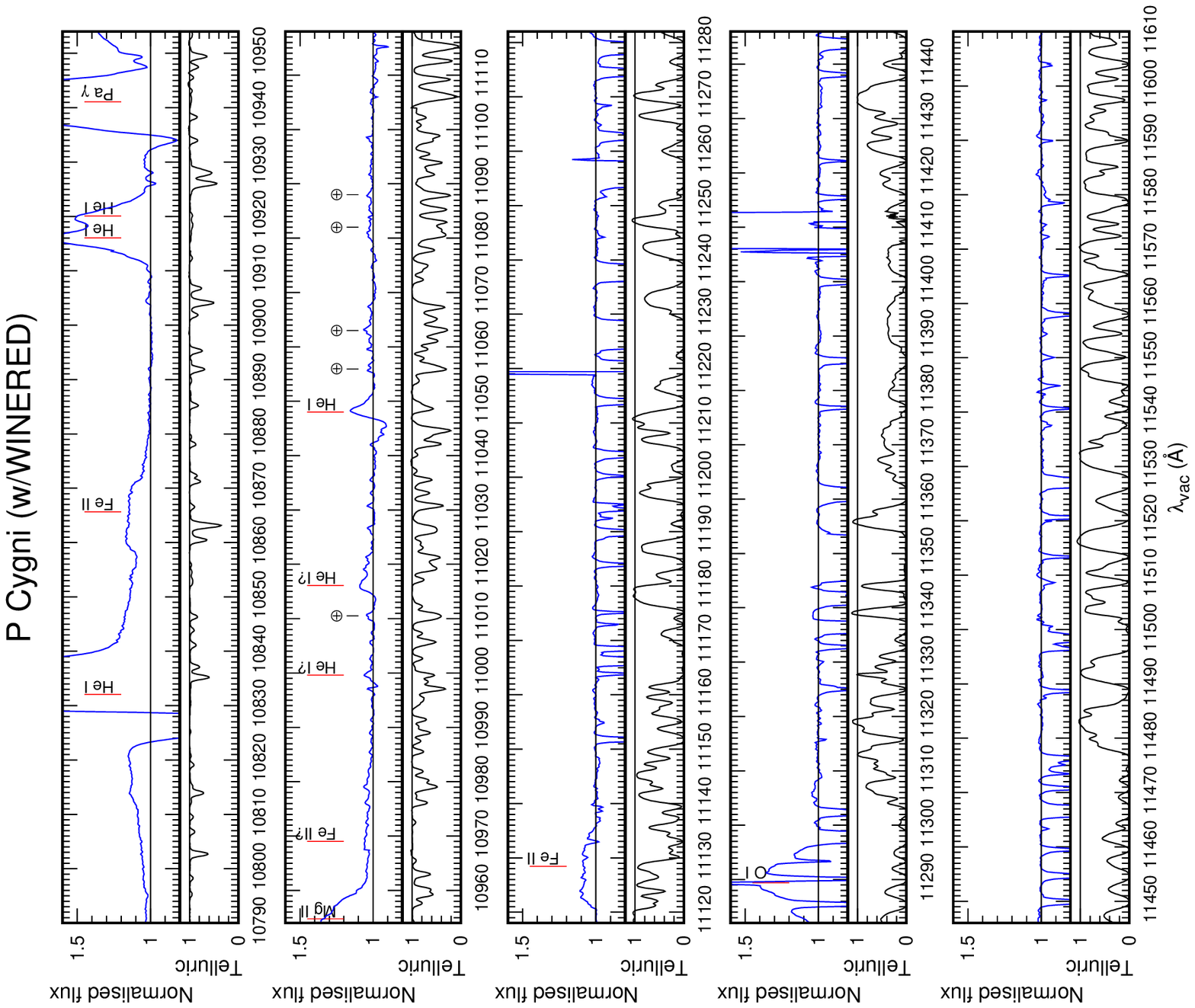}  
\caption{
{\it continued}
}
\label{spectrum2}
\end{figure*}

\addtocounter{figure}{-1}
\begin{figure*}
\centering
\includegraphics[angle=270,width=160mm]{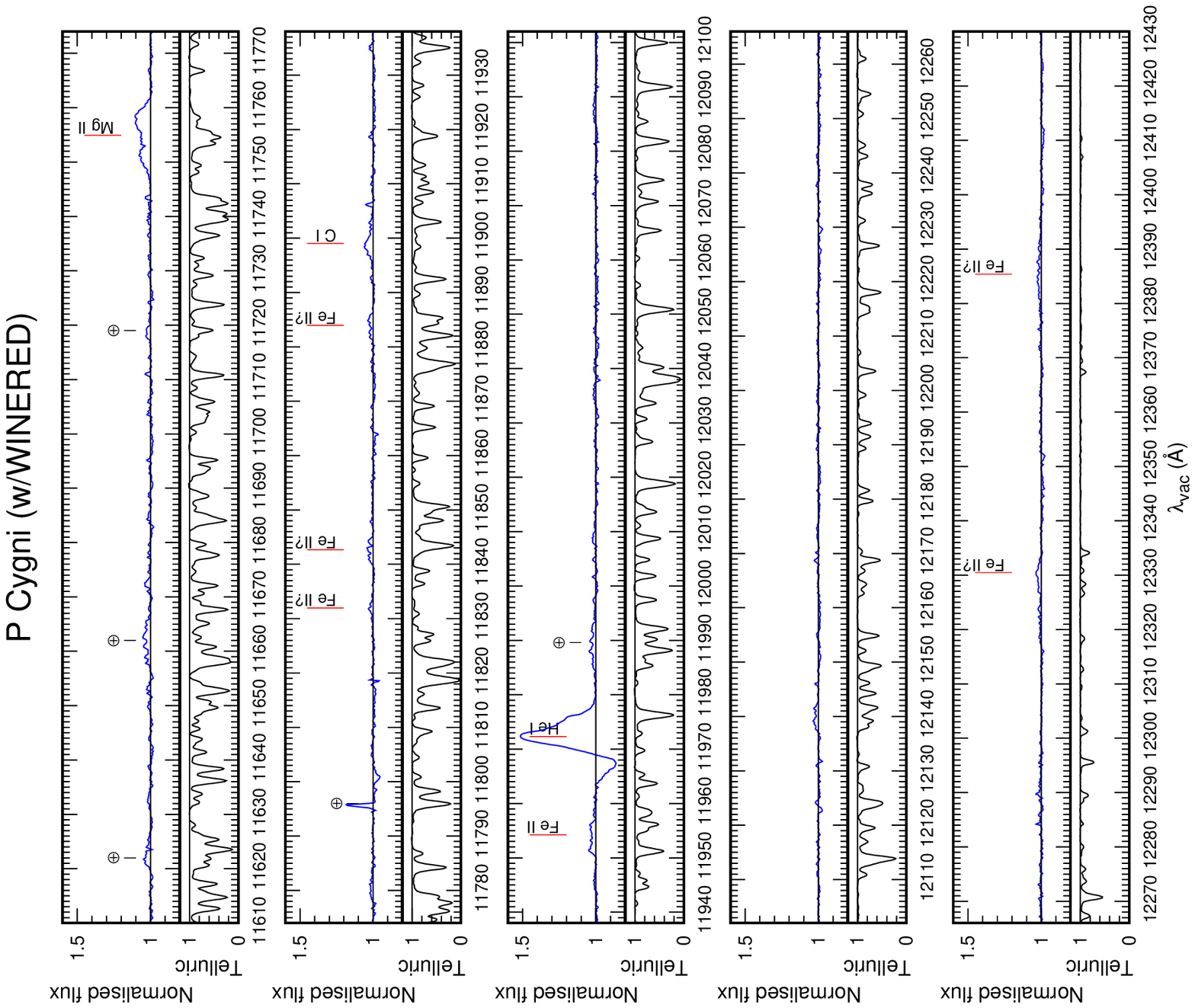}  
\caption{
{\it continued}
}
\label{spectrum2}
\end{figure*}

\addtocounter{figure}{-1}
\begin{figure*}
\centering
\includegraphics[angle=270,width=160mm]{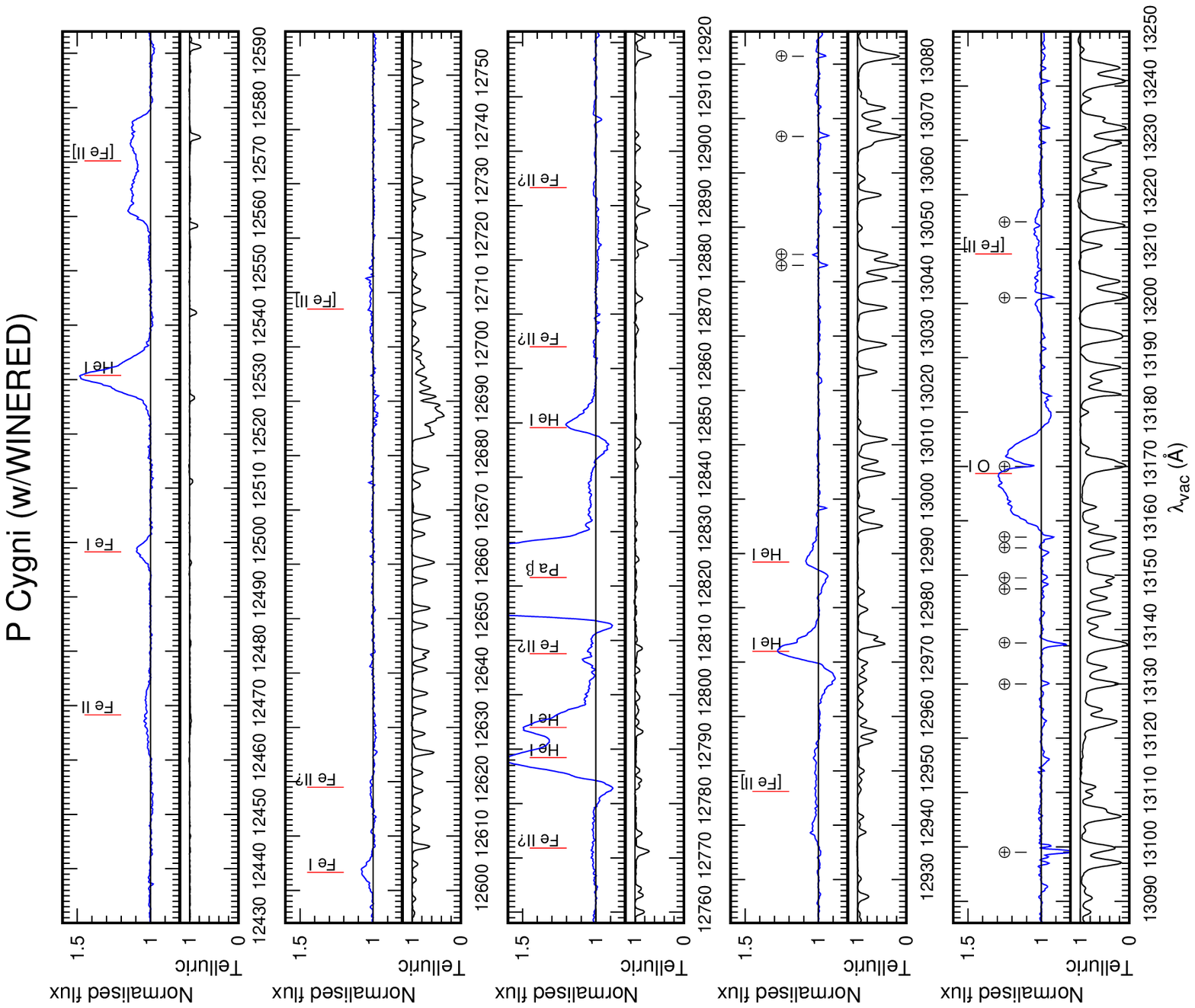}  
\caption{
{\it continued}
}
\label{spectrum2}
\end{figure*}

\begin{table*}
 \begin{center}
 \caption{Line list
 }\label{tab:linelist}
    \begin{threeparttable}
  \begin{tabular}{lllcll | lllcll}
   \hline
\hline
Line & $\lambda_\mathrm{vac}$ (\AA)& $\lambda_\mathrm{air}$ (\AA)& Transition &Rank$^{\ast1}$ &Ref.$^{\ast2}$& Line & $\lambda_\mathrm{vac}$ (\AA)& $\lambda_\mathrm{air}$ (\AA)& Transition &Rank&Ref.\\
\hline
\ion{Mg}{II}& 9220.7799 &	9218.250 	&	$^2S_1-^2P_3$ &B&&	\ion{Fe}{II}& 10549.378 &	10546.488 	&	$e^6D_9-y^6F_{11}$ &A&1\\	
\ion{Pa}{$\zeta$}& 9231.5498 &	9229.017 	&	&A&4& 		\ion{Fe}{II}& 10638.765 &	10635.851 	&	$e^4G_5-^2F_7$ &B&\\
\ion{Mg}{II}& 9246.8019 &	9244.265 	&	$^2S_1-^2P_1$ &B&&		\ion{He}{I}& 10670.898& 10667.975 	&	$^3P_0-^3S_2$ &A&1\\
\ion{Fe}{II}?& 9264.4448 &	9261.903 	&	$^2[3]_7-^4F_5$ &X&&		\ion{O}{I}& 10732.881 &10729.941 	&	$^5P_6-^5S_4$ &B&\\
\ion{Ni}{II}?& 9268.7769 &	9266.234 	&	$^2F_5-^2[3]_7$ &X&&		\ion{He}{I}& 10832.055 &10829.088 	&	$^3S_2-^3P_0$ &A&1,3,4\\
\ion{Fe}{II}& 9299.4709 &	9296.920 	&	$e^4D_5-^4D_5$ &A&4&		\ion{Fe}{II}& 10865.628 &	10862.652 	&	$z^4F_5-b^4G_7$ &A&1,3,4\\
\ion{Fe}{II}?& 9390.509 &	9387.933 	&	$^4D_7-^4F_5$ &X&&		\ion{He}{I}& 10916.022 &10913.033 	&	$^3D_6-^3F_8$ &A&3,4\\
\ion{Fe}{II}& 9408.653 &	9406.072 	&	$u^4F_5-^4G_5$ &A&4&		\ion{He}{I}& 10920.088 &10917.098 	&	$^1D_4-^1F_6$ &A&1\\
\ion{Fe}{II}?& 9415.1861 &	9412.604 	&	$t^2F_5-^2G_7$ &X&&		\ion{Pa}{$\gamma$}& 10941.089 &	10938.093 	&	&A&1,3,4\\ 
\ion{He}{I}& 9466.1383 &	9463.542 	&	$^3S_2-^3P_0$ &A&4&		\ion{Mg}{II}& 10954.778 &	10951.778 	&	$^2D_3-^2P_1$ &A&1\\
\ion{He}{I}& 9519.2437 &	9516.633 	&	$^3P_4-^3D_6$ &A&4&		\ion{Fe}{II}?& 10969.029 &	10966.025 	&	$^4H_{11}-^2H_{11}$ &X&\\
\ion{He}{I}& 9531.8841 &	9529.270 	&	$^1D_4-^1F_6$ &B&&		\ion{He}{I}?& 10999.573 &10996.561	 	&	$^3D_6-^3P_4$ &X&\\
\ion{Pa}{$\epsilon$}& 9548.5917 &	9545.973 	&	&A&4& 		\ion{He}{I}?& 11016.086 &11013.070	 	&	$^1S_0-^1P_2$ &X&\\
\ion{He}{I}& 9606.0522 &	9603.418 	&	$^1S_0-^1P_2$ &B&&		\ion{He}{I}& 11048.028 &11045.003 	&	$^1P_2-^1D_4$ &A&1\\
\ion{He}{I}& 9628.2812 &	 9625.641	&	$^1P_2-^1D_4$ &B&&		\ion{Fe}{II}& 11128.64& 11125.593 	&	$z^4F_3-b^4G_5$ &A&4\\
\ion{C}{I}& 9661.0834 &	9658.434 	&	$^3P_4-^3S_2$ &A&4&		\ion{O}{I}& 11289.408 &	11286.317 	&	$^3P_2-^3D_4$ &A&3,4\\
\ion{He}{I}?& 9684.8466 &9682.191	 	&	$^1P_2-^1S_0$ &X&&		\ion{Mg}{II}& 11754.846 &	11751.629 	&	$^2F_7-^2D_5$ &B&\\
\ion{He}{I}& 9705.5312 &	9702.870 	&	$^3P_0-^3S_2$ &A&4&		\ion{Fe}{II}?& 11831.826 &	11828.588 	&	$^6D_7-^6D_5$ &X&\\
\ion{Fe}{I}?& 9801.9266 & 9799.239	&	$w^5D_6-g^7D_8$ &X&&		\ion{Fe}{II}?& 11842.663 &	11839.422 	&	$^4G_5-^2[2]_3$ &X&\\
\ion{Fe}{II}& 9868.7106 &	9866.005 	&	$^2[1]_3-^2[3]_5$ &B&&		\ion{Fe}{II}?& 11883.959 &	11880.707 	&	$^4H_{11}-^4I_{11}$ &X&\\
\ion{Fe}{I}& 9887.4294 &	9884.719 	&	$^2[2]_3-^2[3]_5$ &B&&		\ion{C}{I}& 11899.006 &11895.750 	&	$^3D_6-^3P_4$ &A&4\\
\ion{Fe}{I}& 9893.3941 &	9890.682 	&	$^2[5]_9-^2[5]_9$ &B&&		\ion{Fe}{II}& 11954.218 &	11950.947 	&	$^2P_3-^2[1]_3$ &B&\\
\ion{Fe}{I}& 9905.3288 &	9902.614 	&	$e^5F_4-^3D_4$ &B&&		\ion{He}{I}& 11972.336& 11969.060 	&	$^3P_4-^3D_6$ &A&3,4\\
\ion{Fe}{II}& 9934.3666 &	9931.643 	&	$f^4G_9-^2[5]_9$ &B&&		\ion{Fe}{II}?& 12329.941 &	12326.568 	&	$^6S_5-^2[3]_5$ &X&\\
\ion{Fe}{II}& 9959.0524 &	9956.323 	&	$z^4F_9-b^4G_9$ &A&4&		\ion{Fe}{II}?& 12385.369 &	12381.981 	&	$^2G_7-^4G_9$ &X&\\
\ion{N}{I}& 9971.2432 &	9968.510 	&	$^4D_5-^4P_3$ &B&&		\ion{Fe}{II}& 12467.78 &	12464.370 	&	$^6P_5-^4S_3$ &A&4\\
\ion{Fe}{I}& 9984.6167 &	9981.880 	&	$n^7D_6-^2[5/2]_6$ &B&&		\ion{Fe}{I}& 12498.742& 12495.323 	&	$h^5D_8-^5F_{10}$ &B&\\
\ion{Fe}{II}& 10000.079 &	9997.338 	&	$^4P_3-^6D_5$ &A&2&		\ion{He}{I}& 12530.778 &12527.350 	&	$^3S_2-^3P_0$ &A&3,4\\
\ion{Al}{II}?& 10020.14 &	10017.394 	&	$^1G_8-^1F_6$ &X&&		\ion{[Fe}{II]}& 12570.206 &	12566.768 	&	$^a6D_9-a^4D_7$ &A&3,4\\
\ion{He}{I}& 10030.465 &	10027.716 	&	$^3D_6-^3F_8$ &A&2,4&		\ion{Fe}{I}& 12603.308 &12599.861 	&	$e^7F_4-^2[7/2]_6$ &B&\\
\ion{He}{I}& 10033.904 &	10031.154 	&	$^1D_4-^1F_6$ &A&2,4&		\ion{Fe}{II}?& 12618.98 &	12615.528 	&	$^4G_7-^2[3]_7$ &X&\\
\ion{Fe}{I}?& 10037.351 &	10034.600 	&	$e^5F_4-^3P_4$ &X&&		\ion{[Fe}{II]}& 12706.91 &	12703.435 	&	$a^6D_1-a^4D_1$ &A&3,4\\
\ion{Pa}{$\delta$}& 10052.128 &	10049.373 	&	&A&2,4& 		\ion{Fe}{II}?& 12771.147 &	12767.654 	&	$^4H_9-^2[5]_11$ &X&\\
\ion{C}{I}& 10067.629 &	10064.870 	&	$^3P_0-^3P_2$ &B&&		\ion{He}{I}& 12788.483 &	12784.985 	&	$^3D_6-^3F_8$ &A (New)&\\
\ion{He}{I}?& 10074.809 &10072.048	 	&	$^3D_6-^3P_4$ &X&&		\ion{He}{I}& 12794.064 &12790.565 	&	$^1D_4-^1F_6$ &A (New)&\\
\ion{Mg}{II}& 10094.862 &	10092.095 	&	$^2F_5-^2G_7$ &B&&		\ion{Fe}{II}?& 12807.544 &	12804.041 	&	$e^4G_5-u^4F_7$ &X&\\
\ion{Al}{II}& 10110.212 &	10107.441 	&	$^3D_4-^3P_2$ &A&2&		\ion{Pa}{$\beta$}& 12821.584 &	12818.077 	&	&A&3,4\\ 
\ion{N}{I}& 10117.413 &	10114.640 	&	$^4D_7-^4F_9$ &A&2,4&		\ion{He}{I}& 12849.449 &	12845.935 	&	$^3P_4-^3S_2$ &A (New)&\\
\ion{He}{II}?& 10126.367 &10123.592	 	&	 &X&&		\ion{Fe}{II}?& 12864.03 &	12860.512 	&	$^2H_9-^2[4]_7$ &X&\\
\ion{He}{I}& 10141.214 &	10138.435 	&	$^1P_2-^1D_4$ &A&2&		\ion{Fe}{II}?& 12893.295 &	12889.769 	&	$^4H_7-^2[4]_9$ &X&\\
\ion{Fe}{II}& 10176.303 &	10173.514 	&	$z^4D_7-b^4G_9$ &A&4&		\ion{[Fe}{II]}& 12946.204 &	12942.664 	&	$a^6D_5-a^4D_5$ &A&3,4\\
\ion{He}{I}& 10314.044 &	10311.218 	&	$^3P_4-^3D_2$ &A&3,4&		\ion{He}{I}& 12971.986 &12968.439 	&	$^1P_2-^1D_4$ &A (New)&\\
\ion{Fe}{II}& 10436.446 &	10433.587 	&	$w^4G_{11}-^4G_9$ &B&&		\ion{He}{I}& 12988.434 &12984.882 	&	$^3D_6-^3P_4$ &A (New)&\\
\ion{Fe}{II}& 10504.38 &	10501.502 	&	$z^4F_7-b^4G_9$ &A&3,4&		\ion{O}{I}& 13168.732 &13165.131 	&	$^3P_0-^3S_2$ &A&3,4\\
\ion{Fe}{II}& 10525.135 &	10522.252 	&	$^2[2]_3-^6D_1$ &B&&		\ion{[Fe}{II]}& 13209.111 &	13205.499 	&	$a^6D_7-a^4D_7$ &A&3,4\\
\hline
  \end{tabular}
\begin{tablenotes}\footnotesize
    \item[$\ast1$] (A): (Newly) identified
 (B): Probably identified
 (X): Detected but not identified; possible candidates are shown.
    \item[$\ast2$] References: 1:\citet{gro07}, 2:\citet{ros01}, 3:SH06, 4:\citet{ham94}. Note that the linelist in [4] is for $\eta$ Carinae. 
    \end{tablenotes}
  \end{threeparttable}
  \end{center}
\end{table*}
%


\bsp	
\label{lastpage}
\end{document}